\documentclass[aip,rsi,reprint]{revtex4-1}
\usepackage{environ}  
\usepackage{upgreek}
\usepackage[colorlinks=true,linkcolor=blue]{hyperref}%
\expandafter\ifx\csname package@font\endcsname\relax\else
 \expandafter\expandafter
 \expandafter\usepackage
 \expandafter\expandafter
 \expandafter{\csname package@font\endcsname}%
\fi
\usepackage{graphicx}
\usepackage{bm}
\usepackage{textcomp}
\usepackage[utf8]{inputenc}
\usepackage[T1]{fontenc}
\usepackage{mathptmx}
\usepackage{amsmath}
\usepackage{anyfontsize}
\usepackage{placeins}
\DeclareUnicodeCharacter{2212}{-}

\begin{document}
\title{A Three-step Model for Optimizing Coil Spacings Inside Cuboid-shaped Magnetic Shields}

\author{Tianhao Liu}
\email{silasliutianhao@gmail.com}
\affiliation{Physikalisch-Technische Bundesanstalt Berlin, 10587 Berlin, Germany}
\affiliation{Department of Electrical Engineering and Automation, Harbin Institute of Technology, 150001 Harbin, China }
\author{Allard Schnabel}%
\affiliation{Physikalisch-Technische Bundesanstalt Berlin, 10587 Berlin, Germany}%
\author{Jens Voigt}%
\affiliation{Physikalisch-Technische Bundesanstalt Berlin, 10587 Berlin, Germany}%
\author{Zhiyin Sun}
\email{23hnhosava@163.com}
\affiliation{Laboratory for Space Environment and Physical Sciences, Harbin Institute of Technology, 150001 Harbin, China}
\author{Liyi Li}%
\affiliation{Department of Electrical Engineering and Automation, Harbin Institute of Technology, 150001 Harbin, China }
\affiliation{Laboratory for Space Environment and Physical Sciences, Harbin Institute of Technology, 150001 Harbin, China}

\date{\today}

\begin{abstract}
	A three-step model for calculating the magnetic field generated by coils inside cuboid-shaped shields like magnetically shielded rooms (MSRs) is presented. The shield is modelled as two parallel plates of infinite width and one tube of infinite height. We propose an improved mirror method which considers the effect of the parallel plates of finite thickness. A reaction factor is introduced to describe the influence of the vertical tube, which is obtained from finite element method (FEM) simulations. By applying the improved mirror method and then multiplying the result with the reaction factor, the magnetic flux density within the shielded volume can be determined in a fast computation. The three-step model is verified both with FEM and measurements of the field of a Helmholtz coil inside an MSR with a superconducting quantum interference device. The model allows a fast optimization of shield-coupled coil spacings compared to repetitive time-consuming FEM calculations. As an example, we optimize the distance between two parallel square coils attached to the MSR walls. Measurements of a coil prototype of 2.75~m in side length show a magnetic field change of 18~pT over the central 5~cm at the field strength of 2.7~\textmu T. This obtained relative field change of 6~ppm is a factor of 5.4 smaller than our previously used Helmholtz coil.
\end{abstract}
   
\maketitle

\section{Introduction}

Extremely uniform magnetic fields with a high temporal stability are essential to various precision measurements, such as electric dipole moment (EDM) measurements \cite{Sakamoto2015,Slutsky2017,PerezGalvan2011,Abe2018} or magnetic field detector calibration \cite{Zikmund2015a,Bronaugh1995,Wang2019}. For the challenging EDM measurements a nonzero magnetic field gradient enhances vibration noise seen by magnetometers \cite{Voigt2013,Yamazaki2009} and shortens the possible measurement time for a single experiment,  thus deteriorating its statistical sensitivity \cite{Clayton2011,Cates1988,Allmendinger2017}. The final unavoidable field gradient also causes systematic uncertainties like geometric phase shift \cite{Afach2015,Pignol2012}, which is the dominant error contribution to the present neutron EDM upper limit \cite{Abel2020}. More details about the influence of the magnetic field gradient can be found in Refs. \onlinecite{Abel2019a,Dadisman2018}.    

The common method of creating a homogeneous magnetic field is to place a coil set into a multi-layer magnetic shield \cite{Hosoya1991,Wyszynski2017,Altarev2014,Liu2020} which serves to reduce external field perturbations. A number of classic coil configurations have been proposed for generating homogeneous magnetic fields in air, such as the Helmholtz coil or the solenoidal coil  \cite{Nouri2013,Wu2019}. However, the high permeability shielding material alters the field distribution and normally worsens the expected uniformity. By optimizing the coil spacings, the ferromagnetic shield can increase the strength and also the uniformity of the magnetic field produced by coils compared to the same setup placed in an unshielded environment \cite{Hanson1965,Liu2020,Hosoya1991,Bidinosti2014a,Liu2020a}.  The result of spacing optimizations depends on the accuracy of the field calculation including the distortion caused by the shielding material. To date the field generated by coils inside a non-spherical magnetic shield can be solved analytically only for some ideal situations, such as assuming an infinite length or an infinite permeability of the shield \cite{Hanson1965,Liu2020,Bidinosti2014a}. The real shield is often of a cuboid shape with no known analytical solution.

For realistic setups, the finite element method (FEM) allows to calculate the magnetic field distribution, but the calculation time required for an accurate description is still quite long for present computers, due to a large length-to-thickness ratio of the shielding material. Sweeping the coil parameters in FEM models to find an optimal solution takes even longer. For this reason, it is expedient to use a reasonable and sufficiently accurate simplification of the field calculation in order to optimize coil spacings in an efficient way.

For a coil in front of an infinitely large and infinitely thick shielding plate, an accepted treatment is substituting the plate by an imaginary coil on the opposite side of the plate as a reflection of the original coil, known as the method of mirror images \cite{inbookTurouski2013,Liu2018}. Pan et al. applied the standard images method to optimize square coils enclosed in a magnetically shielded room (MSR) \cite{Pan2019}. For the case where the thickness of the plates is several mm, the current of the imaginary coil is, however, influenced by the geometry of the setup as well as the position of the observation point, which is not considered in the method of mirror images. The reduced image current due to the finite width of the ferromagnetic plate has been accounted for in the calculation of several specific cases via FEM \cite{Lee2013}. The accuracy was significantly improved, but the effects of the finite thickness were still neglected.

In this paper, we propose a three-step model to calculate the magnetic field generated by coils inside a cuboid-shaped magnetic shield, which is suitable for rapid optimizations of the coil spacings. The proposed three-step model is verified with a complete FEM calculation and a measurement of a 1.6-m-diameter Helmholtz coil placed inside the Berlin Magnetically Shielded Room 2 (BMSR-2) \cite{Bork2001}. Furthermore, the model is applied to optimize the distance between two square coils attached to the walls of BMSR-2 with respect to the field  homogeneity in the central area of the MSR. The three-step model turns out to be 10 times faster than the pure FEM optimization and delivers exactly the same optimal distance. For the newly designed coil pair prototype of 2.75~m in side length with the calculated optimal distance, the measured maximal relative magnetic field change in a 5~cm region around the local extremum near the center of the coil is 6~ppm, which is an improvement by a factor of 5.4 compared to the value of our previously used Helmholtz coil. 

\section{Three-step model description}
\begin{figure}[ b]    
	\centerline{\includegraphics[width=\columnwidth]{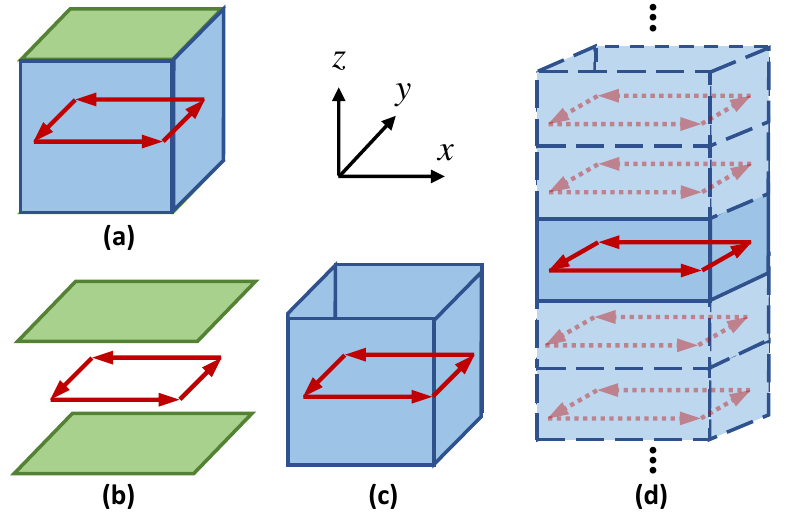}}
	\caption{Development of a mirror model for a cubical shield (a) Schematic view of a coil in a cubic magnetic shield. (b) Coil between infinite parallel plates. (c) Coil in a rectangular tube. (d) Multi-reflection of the coil and the vertical plates.}
	\label{fig:Three-step-model}
\end{figure}

It is assumed that the windings of the coil set are restricted to parallel planes, e.g. Helmholtz coils or Braunbek coils \cite{Abbott2015,Beiranvand2013}. The six faces of a cubic shield as in Fig.~\ref{fig:Three-step-model}(a) are divided into two parts, two horizontal plates parallel to the coil plane and a vertical tube comprising all side walls, as illustrated in Fig.~\ref{fig:Three-step-model}(b) and Fig.~\ref{fig:Three-step-model}(c).

According to the images method, the two horizontal plates of high permeability are treated as mirrors and create multiple mirror images of the coil and also of the vertical tube, thereby extending the tube to a square tube of infinite length as shown in Fig.~\ref{fig:Three-step-model}(d). In our model we assume that the mirror tubes have the same permeability as the original tube, and that the effect of the horizontal plates and of the vertical tube on the field distribution can be considered separately. Therefore, the field generated by a coil inside an MSR of finite thickness and finite permeability can be determined in three steps:

Step 1: Calculate the magnetic flux density $B^{\text{air}}$, generated by the coil in air, with the Biot-Savart law.

Step 2: Calculate the coordinate $\bm{r}_i$ and the current ratio $\tau_{i}^{*}$ of the $i$th mirror coil, which is introduced to replace the effect of the two parallel plates.

Step 3: Calculate the reaction factor curve $\eta_{k}$ in dependence of the position with an FEM simulation once, which is introduced to describe the effect of the surrounding vertical tube in relation to the coil in air.  

We choose a coordinate system with the origin in the center of the magnetic shield, and the position of the original loop is denoted as $\bm{r}_0=(x_0,y_0,z_0)$. The final expression for the $k$ = $x$, $y$ or $z$ component of the magnetic flux density at a position $\bm{r}=(x,y,z)$ inside the enclosure given by the three-step model is 
\begin{equation}
\label{eqn:B_threestep}
	B_{k}^{\text{MSR}}(\bm{r})=\sum_{i=-\infty}^{\infty} B_{k}^{\text {air }}\left(\bm{r}_i^{\prime}\right) \cdot \tau_{k, i}^{*}(x, y, z) \cdot \eta_{k}\left(\bm{r}_i^{\prime}\right) ,
\end{equation} 
where $\bm{r}_i^{\prime}=\bm{r}-\bm{r}_i$. The sign of $i$ indicates the location of the mirror coil, i.e., positive and negative mean above and below the original coil. The sum in Eq.~(\ref{eqn:B_threestep}) is over the original coil $(i = 0)$ and all mirror coils. Without loss of generality, in the later analysis we focus on the $z$ component of the magnetic field since the considered coil is positioned in the $xy$-plane and thus $B_z$ is the most important component regarding field uniformity. The other components can be derived accordingly. For convenience, we also let the axis of the coil be colinear with the $z$ axis, leading $x_0=y_0=0$. 

The principle of the three-step model can be applied to a coil with any shape in one plane. Here, the two most frequently used coil shapes, a circular coil and a square coil, are used as examples to illustrate our model. 

\section{Step 1: coils in air}
The magnetic flux density $B_z$ created by a circular coil loop carrying a current $I$ in air at the observation point \linebreak $\bm{r} (x,y,z)$ with $x^2+y^2<a^2$ is
\begin{equation}
	\label{eqn:Bz_loop_air}	
	B_{z}^{\text{air}}=\frac{\mu_{0}aI}{2} \int_{0}^{+\infty} \zeta J_{1}(\zeta a) J_{0}(\zeta \sqrt{x^{2}+y^{2}}) e^{-|\zeta(z-z_0)|} d \zeta,
\end{equation}
where $\mu_0$ is the permeability of the vacuum, $a$ is the radius of the coil, $z_0$ is the $z$ coordinate of the coil plane and $J_{0(1)}$ is the zeroth (first) order Bessel function of the first kind \cite{inbookCelozzi2008}. Here we use the solution with Bessel functions instead of the more familiar equivalent expression involving elliptic integrals \citep{inbookJackson1998}. We need this solution for the comparison done in section \ref{sec:Mirror}. 

A square coil located in the plane $z = z_0$ is composed out of four identical current segments of equal length $L_\text{coil}$. The center of the $j$th wire segment is denoted as $S_j (x_j, y_j, z_0)$. For the segment with a current in $x$ direction, the magnetic flux density is solved according to the Biot-Savart law as \cite{Hanson2002} 
\begin{equation}
	\label{eqn:Bz_squarex_air}	
	\begin{array}{ll}
	{B_{z}^{j}=\frac{\mu_{0} I_{j}\left(y-y_{j}\right)}{4 \pi \rho^{2}}\left(\frac{\left(x-x_{j}\right)-l_{c}}{\sqrt{\left(x-x_{j}-l_{c}\right)^{2}+\rho^{2}}}-\frac{\left(x-x_{j}\right)+l_{c}}{\sqrt{\left(x-x_{j}+l_{c}\right)^{2}+\rho^{2}}}\right),} \\
	{j=1,3},
	\end{array}
\end{equation}
where $\rho^2=(y-y_j)^2+(z-z_0)^2$ , $l_\text{c}$ = $L_\text{coil} / 2$ and $I_1 = -I_3= I$. The $x$ coordinate and $y$ coordinate in Eq.~(\ref{eqn:Bz_squarex_air}) have to be interchanged for segments carrying current in $y$ direction, leading to
\begin{equation}
	\label{eqn:Bz_squarey_air}	
	\begin{array}{ll}
	{B_{z}^{j}=\frac{\mu_{0} I_{j}\left(x-x_{j}\right)}{4 \pi \rho^{2}}\left(\frac{\left(y-y_{j}\right)-l_{c}}{\sqrt{\left(y-y_{j}-l_{c}\right)^{2}+\rho^{2}}}-\frac{\left(y-y_{j}\right)+l_{c}}{\sqrt{\left(y-y_{j}+l_{c}\right)^{2}+\rho^{2}}}\right),} \\
	{j=2,4},
	\end{array}
\end{equation}
where  $\rho^2=(x-x_j)^2+(z-z_0)^2$ , $I_2 = -I_4= I$. The total magnetic field generated by a square coil is the sum of the contributions of all four wire segments 
\begin{equation}
	\label{eqn:Bz_square_air}	
	B_{z}^{\mathrm{air}}=\sum_{j=1}^{4} B_{z}^{j}.
\end{equation}

\section{Step 2: coils between two infinite plates}

In step 2, the impact of the two parallel plates on the field distribution is analyzed. It is necessary to obtain the $z$ coordinate $z_i$  and current ratio $\tau_i^*$ of the $i$th mirror coil generated by the two parallel plates. 

\subsection{Classical mirror model for parallel plates}
\begin{figure}[ b]    
	\includegraphics[width=0.8\columnwidth]{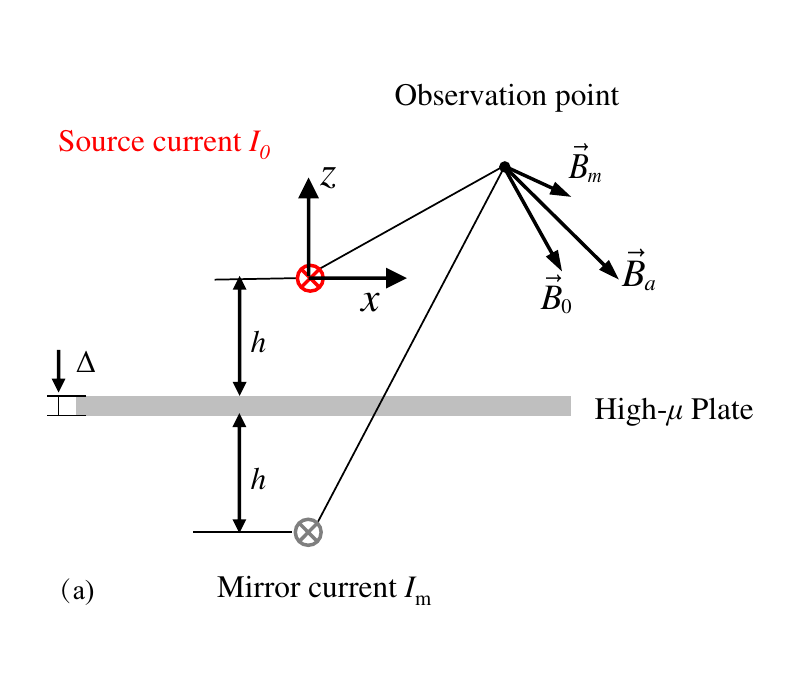}
	\includegraphics[width=0.8\columnwidth]{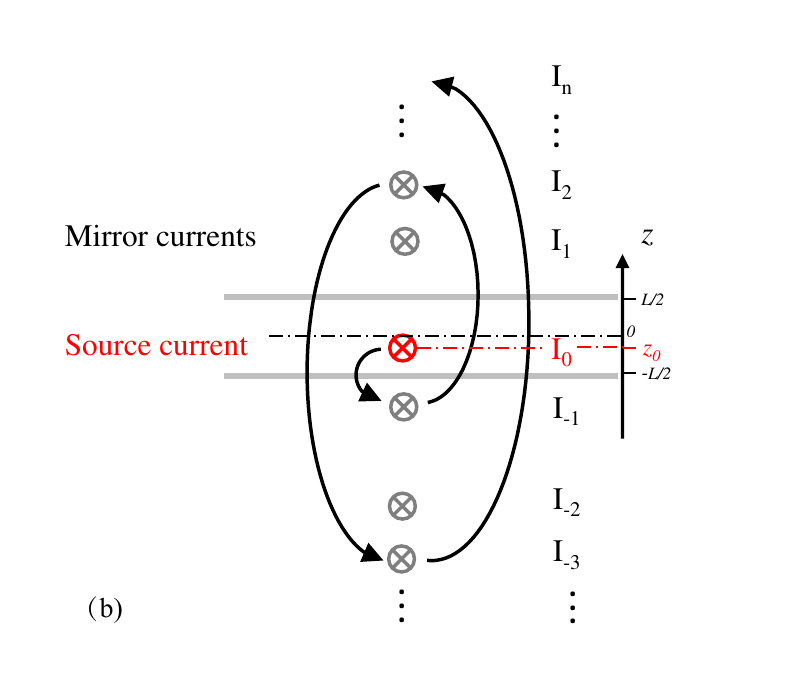} 
	\caption{(a) Classical mirror model for a current $I_0$ in front of a single high-$\mu$ plate of thickness $\Delta$. $h$ is the distance from the current to the plate. The magnetic field $\bm{B}_\text{a}$ on the upper side of the plate is the sum of $\bm{B}_0$, the field generated by the source current, and $\bm{B}_\text{m}$, generated by the mirror current. (b) The extended case with two opposite parallel plates leading to multiple reflections. $L$ is the distance between the two plates and $z_0$ is the $z$ coordinate of the original current. The $i$th mirror current is denoted as $\text{I}_i$.}
	\label{fig:Mirror-model}
\end{figure}

The classical idea of mirror images for a wire in front of one permeable plate is shown in Fig.~\ref{fig:Mirror-model}(a). The effect of the high-permeability plate on magnetic fields can be substituted by the effect of a mirror wire at the same distance to the plate surface as the original wire but on the opposite side. If the thickness $\Delta$ of the plate is large enough, the value of the mirror current $I_\text{m}$ can be written as \cite{inbookTurouski2013}
\begin{equation}
	\label{eqn:I_m}	
	I_\text{m}=\frac{\mu_\text{r}-1}{\mu_\text{r}+1}I_0,
\end{equation}
where $\mu_\text{r}$ is the relative permeability of the plate, and $I_0$ is the source current. The magnetic field $\bm{B}_{\text{a}}$ on the upper side of the plate in Fig.~\ref{fig:Mirror-model}(a) is the sum of $\bm{B}_0$, the field generated by the source current $I_0$ , and $\bm{B}_\text{m}$, generated by the mirror current $I_\text{m}$. 

Mimicking the transportation of light in two opposite mirrors, a wire in between two parallel plates that are treated as mirrors will be repetitively reflected, thus generating multiple mirror wires \cite{Pan2019,Lee2013}. The $i$th mirror wire is labelled as $\text{I}_i$. The absolute value $| i |$ is the number of reflections needed to create this mirror wire. One of the two reflection roads starting from the original wire can be described as: $\text{I}_0 \rightarrow \text{I}_{-1} \rightarrow \text{I}_2 \rightarrow \text{I}_{-3}$, …, and is illustrated in Fig.~\ref{fig:Mirror-model}(b). The second road, not shown, combines the other reflections. The current $I_i$ for the $i$th mirror wire is
\begin{equation}
	\label{eqn:I_i}	
	I_{i}=\left( \frac{\mu_\text{r}-1}{\mu_\text{r}+1}\right) ^{|i|}I_0,
\end{equation}
and the $z$ coordinate of the $i$th mirror wire is 
\begin{equation}
	\label{eqn:z_i}
	z_{i}=(-1)^i*z_{0}+i*L. 
\end{equation}
where $z_{0}$ is the $z$ coordinate of the original wire and $L$ denotes the distance between the two plates. Herein, the origin of the $z$ coordinate is in the middle of the two plates. Eqs.~(\ref{eqn:I_i}) and (\ref{eqn:z_i}) are independent of the shape of the original current path and thus are valid for circular coils, too. The total field is the sum of the contributions from the original wire and all the mirror wires. In general, the multiple mirror method provides an intuitive and easy-to-calculate solution to the considered problem.  However, the accuracy of this model is only guaranteed for infinitely thick plates. For a finite-thickness plate, the value of the mirror current is reduced compared to Eq.~(\ref{eqn:I_m}).

\subsection{Analytical calculation of the mirror current ratio for plates of finite thickness}
\label{sec:Mirror}
To calculate the magnetic field of coils inside a practical MSR, the assumption of an infinite thickness of the shielding material is hard to justify. The material is only a few mm thick. Together with realistic values of $\mu_{\text{r}}$ of 15000 – 30000 (the effective $\mu_{\text{r}}$ for BMSR-2 is 17500 \cite{Bork2001}) the influence of the finite thickness on the field could reach several percent. Therefore, our model includes the influence of the finite thickness.

In multiple mirror models the parallel plates are treated sequentially and each time only one plate is considered as a mirror. Therefore, the analysis with a single plate is applicable to obtain the current value $I_i$ for all mirror wires.

Consider a coil loop of radius $a$ and current $I_0$ located at $ z_{0}=0$ in the $xy$-plane. A plate of thickness $\Delta$ is located at $z = - h$. We choose a Cartesian coordinate system whose origin is placed in the center of the loop. The solution for the magnetic flux density in the region $z>-h$ and $x^2+y^2<a^2$ is solved in Ref. \onlinecite{inbookCelozzi2008} as  
\begin{equation}
\begin{aligned}
	\label{eqn:Bz_plate_loop}
	B_{z}=& \frac{\mu_{0} a I_{0}}{2} \int_{0}^{+\infty} \zeta J_{1}(\zeta a) J_{0}(\zeta \sqrt{x^{2}+y^{2}})e^{-\zeta z} d \zeta  \\
	 +& \frac{\mu_{0} a I_{0}}{2} \int_{0}^{+\infty} T(\zeta) \zeta J_{1}(\zeta a) J_{0}(\zeta \sqrt{x^{2}+y^{2}}) e^{-\zeta|z+2h|} d \zeta \\
	=& B_z^{\text{air}} + B_{z}^{\text{plate}},
\end{aligned}
\end{equation}
with
\begin{equation}
	\label{eqn:T}
	T(\zeta)=\frac{\left(1-\mu_{\text{r}}^{2}\right)\left(1-e^{2\zeta\Delta}\right)}{\left(1+\mu_{\text{r}}\right)^{2} e^{2\zeta\Delta}-\left(1-\mu_{\text{r}}\right)^{2}}.
\end{equation}
The first term $B_{z}^{\text{air}}$ in Eq.~(\ref{eqn:Bz_plate_loop}) is the field generated by the loop in air (see Eq.~(\ref{eqn:Bz_loop_air})) so that the second term is the impact of the permeable plate, noted as $B_{z}^{\text{plate}}$. To interpret the effect of the plate with the mirror method, we have to give $B_{z}^{\text{plate}}$ the form $ \tau B_{z}^{\text{air},z_\text{m}}$, which can be obtained by 
\begin{equation}
	\label{eqn:transform}
	B_{z}^{\text{plate}} = \frac{B_{z}^{\text{plate}}}{B_{z}^{\text{air},z_\text{m}}}B_{z}^{\text{air},z_\text{m}}=\tau B_{z}^{\text{air},z_\text{m}}.
\end{equation}
where $B_{z}^{\text{air},z_\text{m}}$ is the field generated by an imaginary air coil of the same radius $a$ and current $I_0$ as the original coil, but located at $z_\text{m} = - 2h$ (the mirror position of the original coil). $\tau$ is a proportionality factor. Substituting  $B_{z}^{\text{plate}}$ and  $B_z^{\text{air}}$ (Eq.~(\ref{eqn:Bz_plate_loop})) into Eq.~(\ref{eqn:transform}), one obtains
\begin{equation}
\label{eqn:tau}
	\tau= \frac{B_{z}^{\text{plate}}}{B_{z}^{\text{air},z_\text{m}}} = \frac{\int_{0}^{+\infty} T(\zeta) \zeta J_{1}(\zeta a) J_{0}(\zeta \sqrt{x^{2}+y^{2}}) e^{-\zeta\left|z-z_{\mathrm{m}}\right|} d \zeta}{\int_{0}^{+\infty} \zeta J_{1}(\zeta a) J_{0}(\zeta \sqrt{x^{2}+y^{2}}) e^{-\zeta\left|z-z_{\mathrm{m}}\right|} d \zeta}.
\end{equation}

$\tau$ is independent of the source current $I_0$ and it can also be interpreted as the current ratio of the mirror loop to the original loop. It can be deduced from Eq.~(\ref{eqn:tau}) that $\tau<1$, which means that the mirror current value is always smaller than the source current. In the limit $\Delta \rightarrow \infty$, $T$ defined in Eq.~(\ref{eqn:T}) reduces to $T = (\mu_\text{r}-1)/(\mu_\text{r}+1)$, which is independent of the integral variable $\zeta$. Inserting this term into the definition of $\tau$ in Eq.~(\ref{eqn:tau}), one finds that $\tau = T=(\mu_\text{r}-1)/(\mu_\text{r}+1)$, which is identical to the constant mirror current ratio defined in the classical images method (see Eq.~(\ref{eqn:I_m})). 

\begin{figure}[ b]    
	\centerline{\includegraphics[width=\columnwidth]{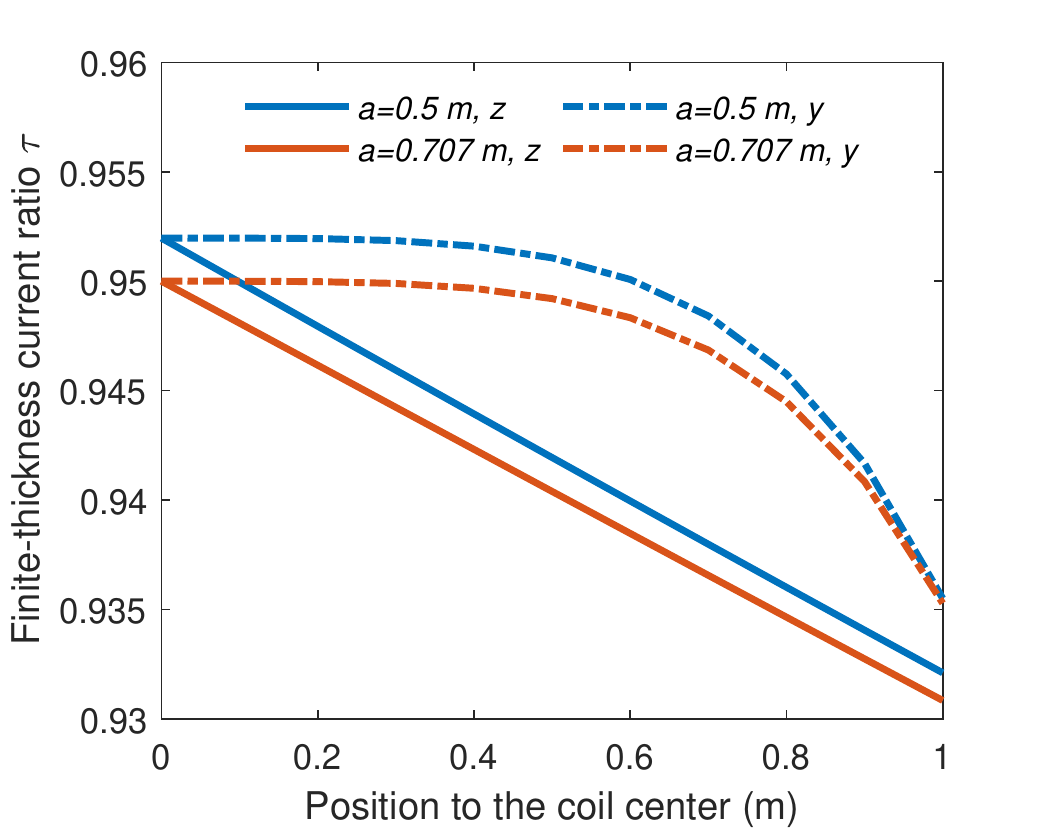}} 
	\caption{The finite-thickness mirror current ratio  of a circular loop of radius $a$ placed in the $xy$-plane with the center at the origin of the coordinate system as a function of the observation position along the $z$-axis (lines) and $y$-axis (dashed lines). The loop is in front of a plate of $\mu_\text{r}$ = $2 \times 10^4$ and $\Delta$ = 2~mm, and located at $z$ = -1~m. }
	\label{fig:Current_ratio}
\end{figure}

After the transformation in Eq.~(\ref{eqn:transform}), the magnetic field can be written as $B_z=B_z^{\text{air}} + \tau B_{z}^{\text{air},z_\text{m}}$, which is an alternative to the analytic expression in Eq.~(\ref{eqn:Bz_plate_loop}). The analytical expression for the magnetic field generated by a coil in front of a plate does not exist for most of coil geometries, such as the square coil loop. In this case, the mirror method can still be applied as an approximate solution via taking the current ratio $\tau$ of a circular loop as that of other coil geometry. Fig.~\ref{fig:Current_ratio} shows $\tau$ as a function of an observation point on the $z$-axis and the $y$-axis. For a plate of $\mu_\text{r} = 20000$ and $\Delta = 2$~mm, $\tau$ is about 5\% smaller than the ratio in the classical mirror method, which is 0.9999. It was found via FEM simulations that $\tau$ changes by less than 1\% if the square loop of side length $L_\text{coil} = 1$~m is approximated by an internal tangent circular loop $(a = 0.5~\text{m})$ or external tangent circular loop $(a = 0.707~\text{m})$. For the further calculation, we always choose the average of these two radii as the radius for an equivalent circular loop, that is $r_\text{equ}=L_\text{coil} (1+\sqrt{2})/4$ substituting $a$ in Eq.~(\ref{eqn:tau}). 

The current ratio defined in Eq.~(\ref{eqn:tau}) for a plate of finite thickness can be also applied to the multiple mirror method, which is used to model a coil in between two parallel plates. For the $i$th mirror loops, which are created by reflecting the original loop  $| i |$ times, the overall current ratios are the successive product of $| i |$ independent current ratios as
\begin{equation}
	\label{eqn:tau_star}
	\tau_{i}^{*}(x, y, z)=\left\{\begin{array}{ll}
	{\displaystyle \prod_{n=1}^{|i|} \tau_{n(-1)^{i+n}}(x, y, z)} & {i>0,} \\
	{\displaystyle \prod_{n=1}^{|i|} \tau_{n(-1)^{i+n-1}}(x, y, z)} & {i<0,}
	\end{array}\right.
\end{equation}
where the $z$ coordinate of the $n$th mirror loop $z_n$ is given in Eq.~(\ref{eqn:z_i}) and  $\tau_{i}$ is the current ratio for a coil located at $z_i$ in front of a single plate, as defined in Eq.~(\ref{eqn:tau}). Compared to the classical mirror current ratio in Eq.~(\ref{eqn:I_i}), $\tau_{i}^{*}$ takes the finite-thickness of plates into consideration, thus named as finite-thickness current ratio.The magnetic flux density at the observation point $(x, y, z)$ between the two parallel plates considering 2$M$ mirror loops is 
\begin{equation}
	\label{eqn:Bz_square_double_plates}	
	B_{z}=\sum_{i=-M}^{M} \tau_{i}^{*}\left(x, y, z \right)B_{z}^{\text {air }}\left(x, y, z-z_i\right),
\end{equation}
where $i = 0$ for the original loop and $\tau_{0}^{*}=1$. For a mirror loop with an increasing order $i$ the current ratio  $\tau_{i}^{*}$  decreases while the distance to the observation point $| z-z_i |$  increases, making its contributions to the total magnetic field in Eq.~(\ref{eqn:Bz_square_double_plates}) decrease rapidly. For a certain accuracy, this infinite sum can be truncated and the resulting numerical error will be discussed in section \ref{sec:Veri}.    

\section{Step 3: coils inside a vertical tube}
The effect of the high-permeability infinite tube is quantitatively characterized by a reaction factor $\eta$ , which is a function of the observer position $(x, y, z)$ inside the tube and can be written as 
\begin{equation}
	\label{eqn:eta}	
	\eta(x, y, z)=B^{\mathrm{s}}(x, y, z) / B^{\mathrm{air}}(x, y, z),
\end{equation}
where $B^\text{s}$ and $B^\text{air}$ are the $z$ component of the magnetic flux density generated by the coil with the tube and without the tube (in air). $\eta > 1$ means the magnetic field strength is augmented by the high-permeability tube at the observation point. The definition of $\eta $ is independent from the location of the coil. Due to the difficulties in analytically determining $B^\text{s}$, we applied FEM simulations in COMSOL v5.4 to obtain $\eta$. In the FEM model, the height of the tube is set 10 times larger than its width to approximate the infinite length. The shielding tube is modelled as a surface assigned with the magnetic shielding boundary condition with a certain thickness $\Delta$. The usage of this boundary condition avoids the meshing in the direction of the thickness of the material which is a problem due to the large length-to-thickness ratio. By assuming a coil located in the $xy$-plane at the position $z_0$=0,  $\eta $  can be calculated faster using the symmetries ($\eta (x,y,z_0)= \eta (-x,y,z_0) = \eta (x,-y,z_0) = \eta (x,y,-z_0)$) with a one-eighth model in COMSOL as described in Ref. \onlinecite{WalterFrei2014}. For a coil located at any other $z_0$ position, $\eta$ can be obtained with a coordinate transformation.    

\begin{figure}[ b]    
	\centerline{\includegraphics[width=\columnwidth]{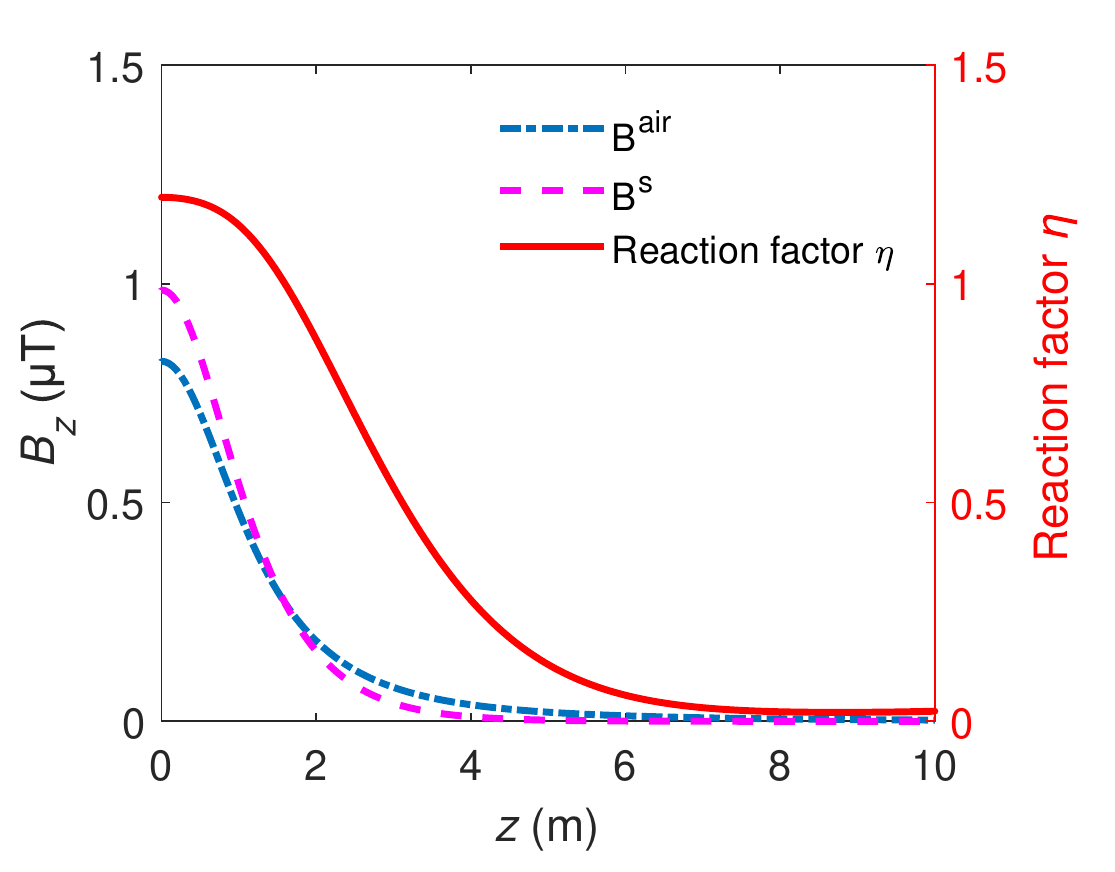}} 
	\caption{Magnetic field density and reaction factor $\eta$ of a square coil located in the $xy$-plane at $z_0=0$, carrying a current of $I=2$~A inside a high-$\mu$ rectangular tube, along the $z$-axis (x=y=0) under the conditions of $\mu_\text{r}$ = $3 \times 10^4$ and $\Delta$ = 4~mm. The side length of the square coil is 2.75~m and that of the rectangular tube is 3.2~m. }
	\label{fig:Eta}
\end{figure}

The FEM result for a square coil in the $xy$-plane at $z_0=0$ with 2.75~m side length inside a high-$\mu$ rectangular tube of 3.2~m side length is plotted in Fig.~\ref{fig:Eta} as a function of the observer position. The left axis shows  $B^\text{air}$ and $B^\text{s}$ while the right axis shows $\eta$ . The same current produces a larger field in the coil center when the coil is inside a shield ( $\eta > 1$), as expected due to the reduction of the magnetic resonance. For $z > 2$~m, $\eta$ becomes less than 1 because the surrounding high-permeability tube attracts most of the flux lines, leading to a strong decrease between 2~m and 6~m. For $z > 6~\text{m}$, $\eta$ is already less than 0.1. 

Once $\eta$ is calculated by FEM for enough sample points, $\eta$ can be interpolated to any observation point without using FEM again. Since the tube is infinitely long, any coil of the same shape in the $xy$-plane obeys the same $\eta$ behavior when the parameter $z$ of the $\eta$ curve is replaced by relative distance $z^\prime$, the distance between $z$ of the observation point and the $z$ coordinate of the coil center. This means that for calculating the field from the original coil and the mirror coils, we only need to calculate $\eta$ once by FEM. This is also true for a coil set with identical windings on different $z$ planes, like a Helmholtz coil. For different coil shapes the $\eta$ curve must be recalculated.

Multiplying the magnetic field generated by each coil in Eq.~(\ref{eqn:Bz_square_double_plates}) with the corresponding reaction factor $\eta$ yields the final expression for the magnetic flux density $B_z$ within the shielded volume observed at $(x,y,z)$ 
\begin{equation}
	\label{eqn:Bz_threestep}
	B_{z}^{\text{MSR}}=\sum_{i=-\infty}^{\infty} B_{z}^{\text {air }}\left(x, y, z_i^{\prime}\right) \cdot \tau_{i}^{*}(x, y, z) \cdot \eta\left(x, y, z_i^{\prime}\right),
\end{equation}
where $z_i^{\prime}=z-z_i$ . Note that the expression for the reaction factor $\eta$ and the current ratio $\tau_{i}^{*}$ are given here only for the $B_z$ component. For the $B_x$ and $B_y$ components, they must be calculated accordingly.

\section{Verification}
\label{sec:Veri}
\subsection{Comparison to FEM calculations}

\begin{figure}[ t]    
	\centerline{\includegraphics[width=\columnwidth]{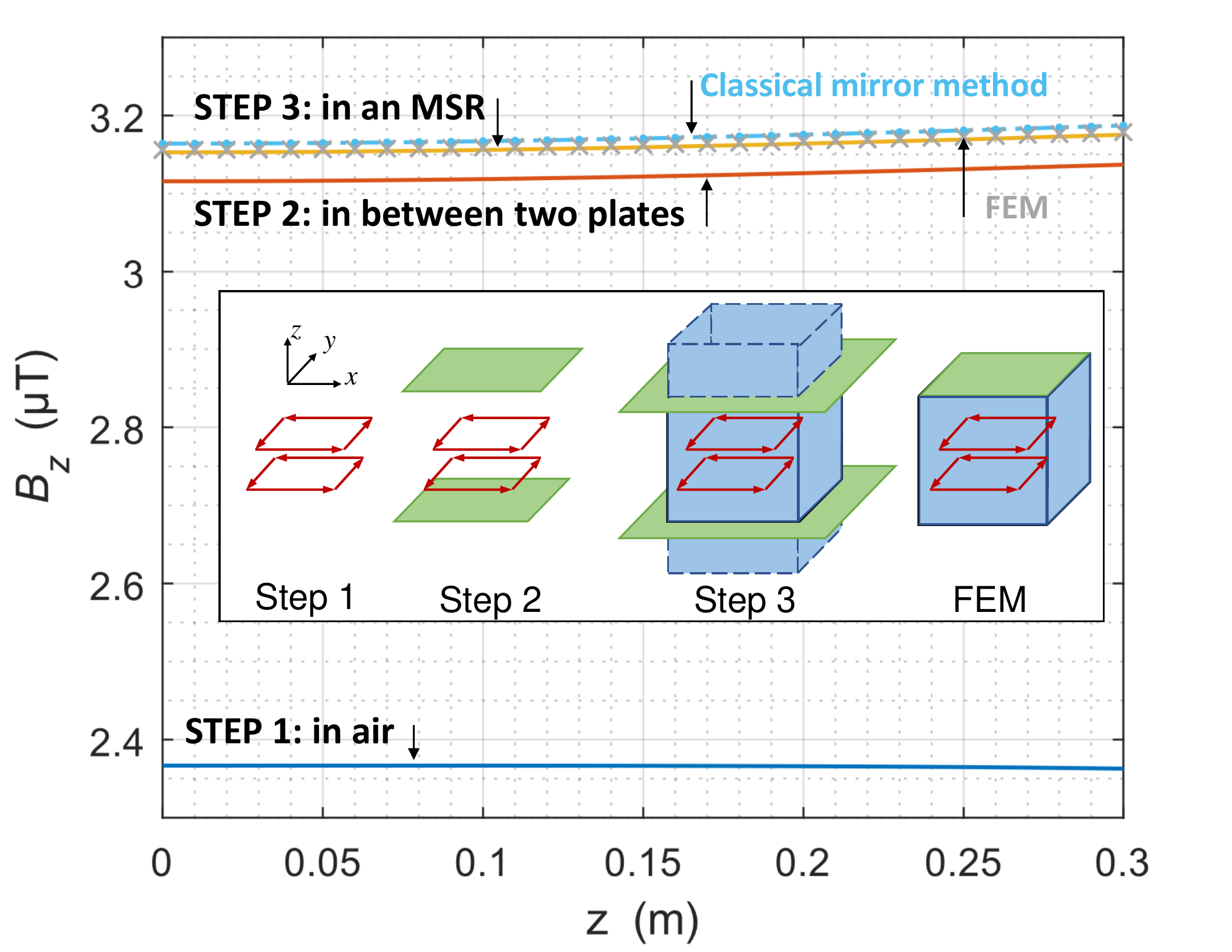}} 
	\caption{Illustration of the three-step model for two square coils inside an MSR under the conditions: $\mu_\text{r}$~=~$3 \times 10^4$, $\Delta$~=~4~mm, $ L_\text{MSR}$~=~3.2~m, $ L_\text{coil} $~=~2.75~m, $d_\text{coil} = 0.5445  L_\text{coil}$  and $I$~=~4~A. The solid lines are the results after each step of the three-step model. The crosses are the FEM results and the dot-dashed curve is the result of three-step model using classical mirror current ratios. Inset: Schematic view of the setup for each step and the FEM calculation.}
	\label{fig:Square}
\end{figure}

Here we demonstrate the three-step model for the case of two square coil loops of side length $L_\text{coil}$ = 2.75~m enclosed in a cubic single-layer MSR of side length $L_\text{MSR}$ = 3.2~m,  a thickness $\Delta$~=~4~mm and a constant $\mu_{\text{r}}$~=~30000. The square coils are in the $xy$-plane, as illustrated in the inset of Fig.~\ref{fig:Square}. The distance between the two coils is set as $d_\text{coil}$ = $0.5445  L_\text{coil}$ to compose a square Helmholtz coil in air \cite{Firester1966}. 

\begin{figure}[t]    
	\centerline{\includegraphics[width=\columnwidth]{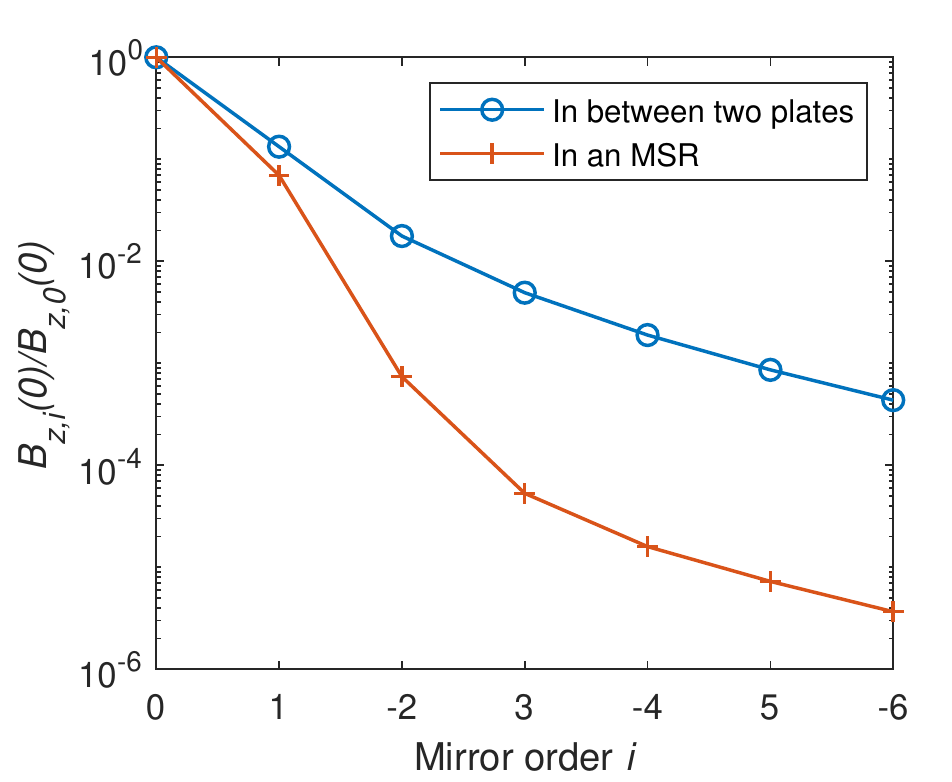}} 
	\caption{The ratio of the magnetic field at the origin ($z=0$) generated by mirror loops to the field generated by the original two coils in an MSR $(i = 0)$. As a comparison, the ratio for coils in between two plates is plotted. Parameters for the coils and the MSR are the same as used in Fig.~\ref{fig:Square}. }
	\label{fig:Field_ratio}
\end{figure}

Fig.~\ref{fig:Square} shows $B_z$ for the single steps of the three-step model and the result for the respective FEM calculation in dependence of the $z$-coordinate, i.e. moving along the coil central axis, where the magnetic field $B_z$ of the coil has the strongest change. The infinite sum over all contributing images in Eq.~(\ref{eqn:Bz_threestep}) was truncated at $i=5$, i.e. we stopped after the fifth mirror image on both sides. The red curve in Fig.~\ref{fig:Field_ratio} shows that the contribution of the next term ($i=6$) for this coil arrangement is already less than 0.05\% for $z =0$~m. We checked that even for $z$ up to 0.3~m, the difference to the $z=0$ curve is less than 5\%. The FEM calculation (done with COMSOL v5.4) is regarded as the benchmark, even though it has a finite uncertainty because it is a mathematical approximation which depends on the chosen boundary condition and the mesh size. Using optimized configurations, the uncertainty is expected to be < 0.1\%. 

Step 1 is the exact solution of the coil in air (Eq.~(\ref{eqn:Bz_square_air})), which underestimates the field inside the shield by 33.2\%. Adding the influence of the two walls parallel to the coils as infinite plates with finite thickness leads to a field 1.2\% below the FEM result. The complete three-step model follows well the increasing FEM trend for larger $z$ values and underestimates the magnetic field by only about 0.1\%, which is already close to the expected uncertainty of the FEM used as benchmark. To demonstrate the influence of introducing the finite-thickness current ratio (Eq.~(\ref{eqn:tau_star})) instead of classical current ratios (Eq.~(\ref{eqn:I_i})) in step 2, we also calculated this case, marked with classical mirror method in Fig.~\ref{fig:Square}. The assumption of an infinite thickness leads to an expected field overestimation. The difference of 0.3\% to the FEM benchmark is small for this case but increases significantly when $\Delta$ or $\mu_{\text{r}}$ are smaller. For example, for the same setup with $\mu_{\text{r}}=$ 10000 and 2000, the differences are 1.0\% and 6.1\%. 

\begin{table}
    \centering
    \caption{The relative error of the field derived from the three-step model compared to the FEM results at the origin with varied parameters values. The setup is the same as in Fig.~\ref{fig:Square} with default values $\mu_{\text{r}}=30000$, $d_{\text{coil}}=1.5$~m and  $L_{\text{MSR}}=3.2$~m. In each column, only one parameter value is altered and the others are left at their default value. Since we would like the coils to be attached to inside walls of the shield, the length of the coil is changed when the length of the MSR is changed according to $L_{\text{coil}}=0.86L_{\text{MSR}}$.}
    \begin{tabular}{c c | c c | c c}
		\hline
    $\mu_{\text{r}}$ & Error & $d_{\text{coil}}$ (m) & Error & $L_{\text{MSR}}$ (m)  & Error \\
      	\hline
    1      & -0.04\%  & 0.5  & -0.05\%  & 2   &  -0.04\%  \\
    50     & 0.94\%   & 1    & -0.07\%  & 2.4 &  -0.06\% \\
    2000   & 1.28\%   & 1.5  & -0.10\%  & 2.8 &  -0.07\% \\    
    10000  & 0.01\%   & 2    & -0.15\%  & 3.2 &  -0.10\% \\  
    30000  & -0.10\%  & 2.5  & -0.27\%  & 3.6 &  -0.10\% \\   
    INF    & 0.00\%   & 3    & -0.32\%  & 4   &  -0.11\% \\
	\hline
	\end{tabular}
    \label{tab:error}
\end{table}

The final deviation of the three-step model is a sum of various effects. Effects leading to an overestimation are: infinite width assumption for the parallel plates, infinite height assumption for the side tube and neglected coupling between the parallel plates and the side tube. In contrast truncation of the sum over all contributions leads to an underestimation. The use of a circular coil as an approximation for the square coil to calculate $\tau^{*}$ and the error of the reaction factor $\eta$ derived via FEM simulations may lead to a positive or a negative error depending on the given geometry. The relative error of the three-step model compared to the FEM results over wide ranges of parameter values for the same setup as in Fig.~\ref{fig:Square} is calculated and summarized in Table \ref{tab:error}. The error never exceeds 1.28\% for all considered cases. $\mu_{\text{r}}$ has the greatest influence, but for a realistic value of $\mu_{\text{r}}=30000$ (permalloy) the error is less than 0.32\% if only the coil distance $d_{\text{coil}}$ or the length of the MSR $L_{\text{MSR}}$ is varied. The maximal influence was found for $\mu_{\text{r}}=2000$ which is a suitable value for ferrites.

\subsection{Comparison to magnetic field measurements }
A measurement to further validate the three-step model was conducted. A movable circular 3-axes Helmholtz coil was used to generate a uniform magnetic field inside BMSR-2. BMSR-2 has a passive shielding factor greater than $7 \times 10^5$ at 0.01 Hz and an effective relative permeability of 17500 \cite{Bork2001}. The innermost 4-mm permalloy layer surrounds a volume of $3.2 \times 3.2 \times 3.216$ $\text{m}^3$. For the three-step model and FEM calculations we included only the innermost shielding layer and neglected all others because the field enhancing effect should be dominated by the innermost layer. Due to this assumption and in order to take care of the additional field increase caused by the remanence of the shielding material, which is not included in the model, we set $\mu_\text{r}$ = 30000.

\begin{figure*}[hbt]    
	\includegraphics[width=\textwidth]{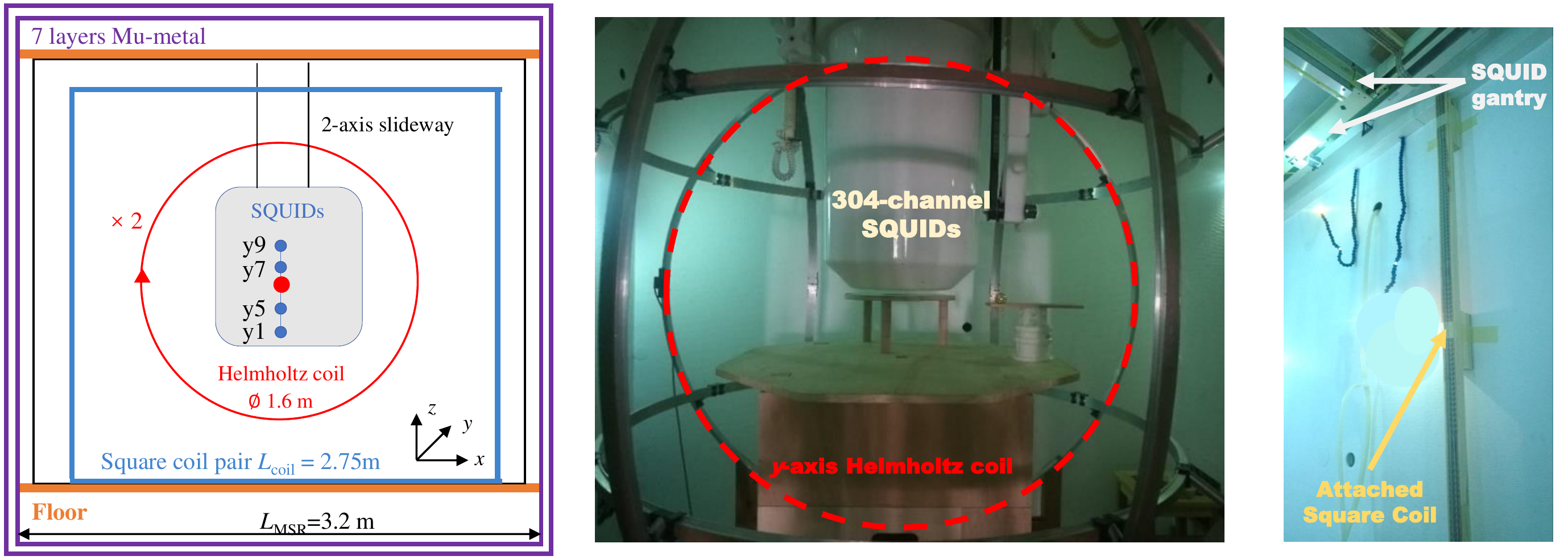}  \\
	\raggedright{ \hspace{25mm} (a) \hspace{70mm} (b) \hspace{55mm} (c)} \\
	\caption[width=\textwidth]{Experimental setup in BMSR-2. (a) Schematic drawing. (b) 3-axes Helmholtz coil with the 304-channel SQUID system inside.The $y$-axis coil used here is marked by a red dashed line. The other coil axes acted as a supporting frame. (c) One coil of the square coil pair prototype attached to the inside walls of BMSR-2.}
	\label{fig:Setup}
\end{figure*}
The Helmholtz coil used was oriented along the $y$-axis of BMSR-2 and has a diameter of 1.6~m. The setup is shown in Fig.~\ref{fig:Setup}(a) and (b). The coil was powered by a low-noise current source from Magnicon \cite{Magnicon_CSE}. The current was measured to be 38.6 mA for each coil with 60 turns in series, generating a 2.6 \textmu T magnetic field in the center measured in a previous nuclear magnetic resonance experiment using $^{3}\text{He}$ spin polarizations. The magnetic flux densities were measured with a 304-channel vector Superconducting Quantum Interference Device (SQUID) system, mounted in a two-axis ball-bearing slideway. The overall absolute positioning error of the SQUID system is $\pm$1 cm. The possible movement range of the SQUID system along the $y$ axis inside the 3-axes Helmholtz coil is from -4 cm to 5 cm. The position of the SQUIDs was moved in steps of 1.0~$\pm$~0.1~cm. The SQUID’s position was locked by a pneumatic brake after each movement. In each position, we measured the magnetic field for 10 seconds to average the noise. Although there are numerous available SQUID channels at different locations, we only used the SQUID channel with the minimum distance to the coil center. In our case, $y$7 was used.

\begin{figure}[ t]    
	\centering
		\includegraphics[width=\columnwidth]{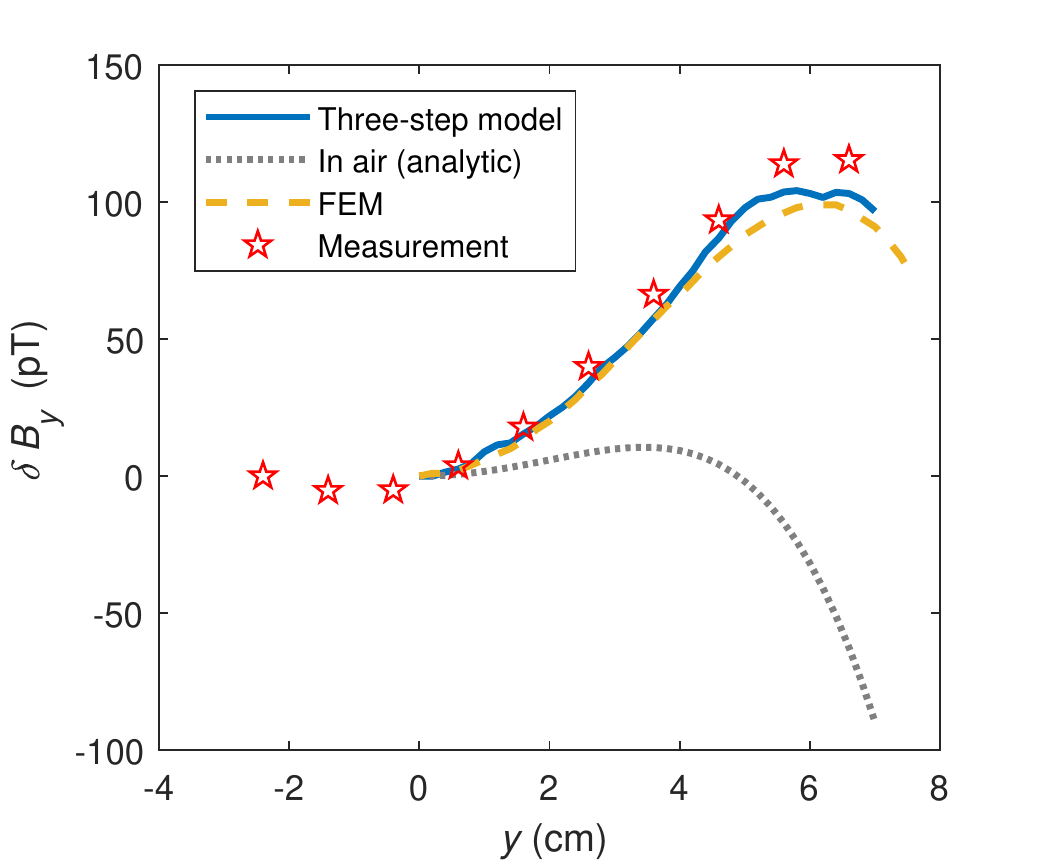}
		\caption{The change of the magnetic field generated by a circular Helmholtz coil of 1.6 m in diameter in BMSR-2 measured with the SQUID $y$7, whose coordinates are $z$ = 2~cm, $y$ = 1.6~cm, $x$ = -2~cm. This field was calculated with the three-step model (blue line) as well as the FEM simulation (yellow dashed line). The field generated by this Helmholtz coil in air was also calculated (grey dotted line). The red stars denote the measured values.}
		\label{fig:Helmholtz}
\end{figure}

In the beginning of the field measurements, the $y$7 SQUID sensor was aligned with the coils axis ($y$-axis) by tilting and rotating the SQUID system to detect the maximum field signal. The remanent field of BMSR-2 in the nT range was measured after degaussing without the coil field and subtracted in data post-processing. Since the DC offset of the SQUID system is unknown, only the change of the magnetic field is plotted in Fig.~\ref{fig:Helmholtz}. The change of the created magnetic field from the field minimum (at -0.4~cm) to a position 5~cm further (at 4.6~cm) is around 98~pT, corresponding to a relative change of 38~ppm. The calculated field change of this Helmholtz coil in air is more uniform, shown as a grey dotted line in Fig.~\ref{fig:Helmholtz}. This demonstrates how the surrounding high permeability walls distort the magnetic field. The results calculated with the three-step model and with the FEM are in acceptable agreement with the measured values. The slight difference may result from the uncertainty of the position of the SQUID chip, and the fact that $\mu_\text{r}$ is not a constant \cite{Andalib2017} as was assumed both for FEM calculations and for the three-step model. Nevertheless, both models successfully predict the maximum of the field at $y$ = 6~cm caused by the coupling to the shielding material. The reason for the This convinced us that the three-step model could be used for the design and optimization of shield-coupled coils. A successful practical example is presented in the next section.

\section{Applications}
Some of the authors have been using BMSR-2 within a collaboration to measure the $^{129}\text{Xe}$ EDM using the free spin precession of Xe gas \cite{Sachdeva2019}. This experiment has stringent requirements on the uniformity of the magnetic field in the region of the sample cell with a size of about 5~cm~×~5~cm~×~5~cm. According to the requirement of the EDM experiment, we defined the magnetic field uniformity in the region of interest as
\begin{equation}
	\label{eqn:u}
	U=\frac{\overline{B_{y}}(y)}{\sigma\left(B_{y}(y)\right)},
\end{equation}
where $\overline{B_{y}}(y)$ and $\sigma\left(B_{y}(y)\right)$ are the average value and the standard deviation of the magnetic field series $B_y(y)$. Here we used a range for $y$ from 0 (coil center) to 5 cm in steps of 1 cm, meaning 6 points in $B_y(y)$. $U$ is a unitless quantity independent of the absolute field strength in the center. 

We built two large square $y$-axis coils. They were attached to the MSR walls as shown in Fig.~\ref{fig:Setup}(c) in order to test if the coils could be permanently hidden behind the internal lining in future. This kind of built-in coil could provide an improved magnetic field homogeneity even though they are located at only a few cm distance to the shielding material. Compared to the Helmholtz coil used before, this built-in coil set would expand the applicable space of the chamber and save time spent in assembling and disassembling the Helmholtz coil system every time measurements with additional magnetic fields are performed. The side length of the square coils was $L_\text{coil}$ = 2.75 m and the distance $d$ between the two coils is the parameter to be optimized. A single coil consists of 8 turns in series, and in total the coil set had a resistance of 18.8 $\Omega$.

\begin{figure}[ t]    
	\includegraphics[width=0.8\columnwidth]{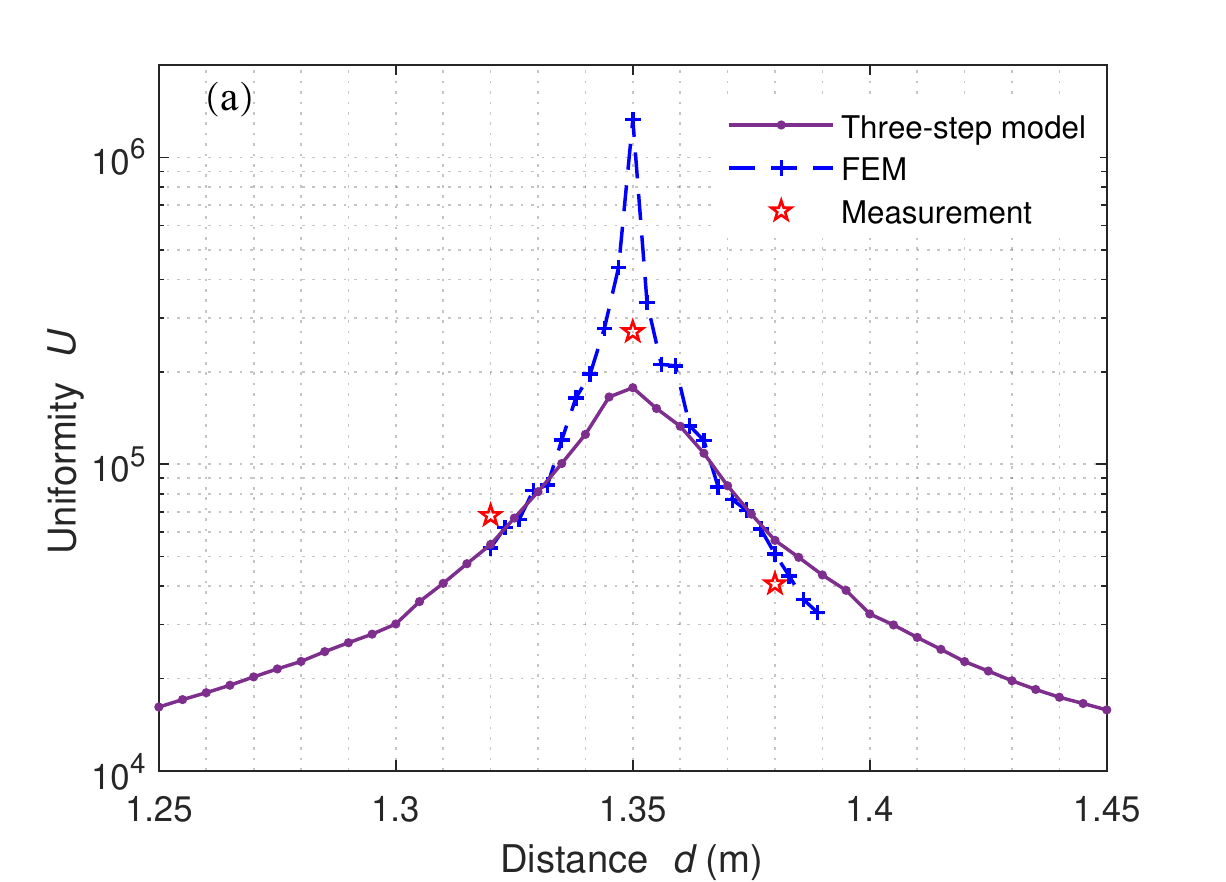}  
	\includegraphics[width=0.8\columnwidth]{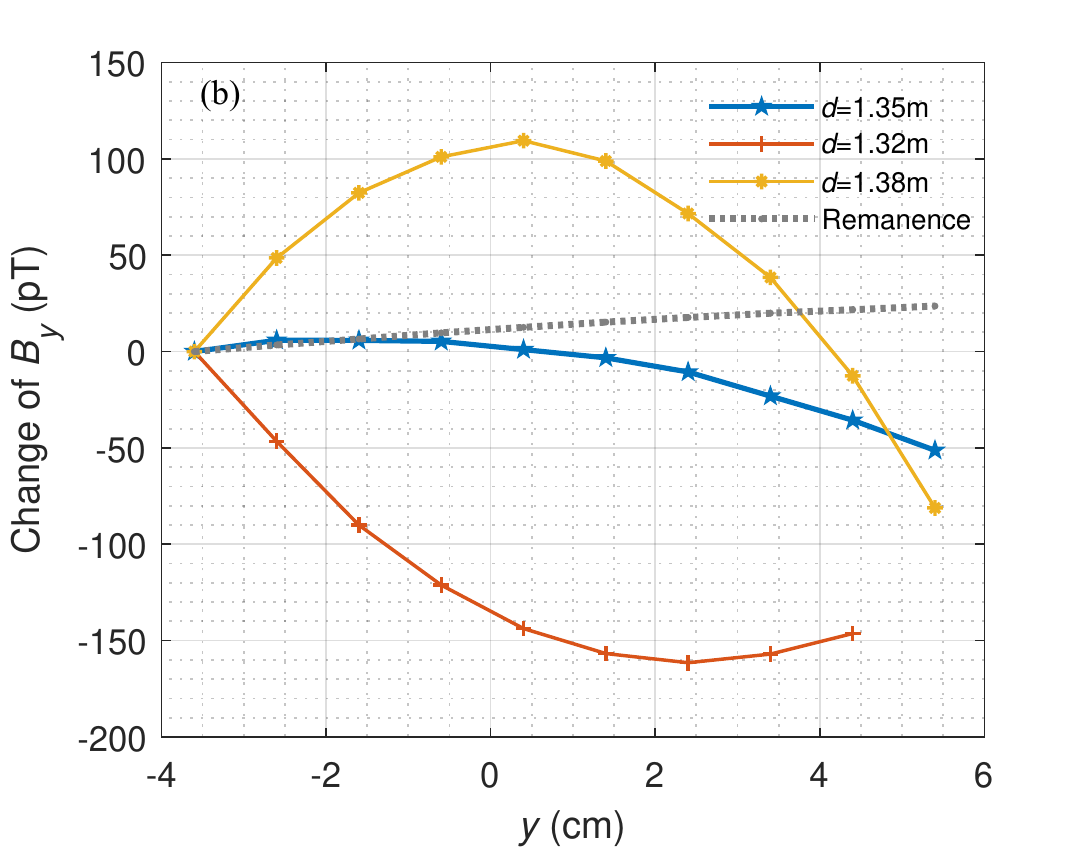}
	\caption{Comparison of the three-step model, FEM simulation and experimental results. (a) The uniformity $U$ as a function of the distance $d$. (b) The measured magnetic field change along the $y$-axis for three $d$ values and the remanent magnetic field. The field amplitude in the center generated by the coil pair with $d$ =1.35 m is 2.7 \textmu T.}
	\label{fig:Opti}
\end{figure}

In order to find the optimal distance $d$ of the coil we first applied the three-step method to obtain the uniformity $U$ as a function of the distance $d$ in the range of 1.25 m to 1.45 m, shown as the light blue curve in Fig.~\ref{fig:Opti}(a). Within 15 minutes of CPU time, we obtained an optimal value of $d$ = 1.35 m. A following FEM simulation was carried out in a narrower range (1.32 m, 1.39 m), shown as the dark blue dashed curve in Fig.~\ref{fig:Opti}(a). After 20 hours of CPU time the optimal distance using repetitive FEM calculations differs by less than 1~mm from the value acquired by the three-step model. The three-step model is not able to reproduce the sharp resonance-like increase of $U$ obtained from the FEM calculation. The main reason is that the reaction factor curve $\eta$ in such a small region of 5 cm is uncertain as $\eta$ is obtained in a FEM model with a tube height over 30 m, leading to a coarse mesh. This uncertainty is also responsible for the unsmooth curve obtained by the three-step model in Fig.~\ref{fig:Helmholtz}. Fig.~\ref{fig:Opti}(a) shows that the position of the coil is critical and should be adjustable around the optimum position. 

The coil pair for the experimental confirmation was set up with a distance of $d$ = 1.35 m ± 0.02 m. A current of 0.42~A through the coil pair generated a magnetic field of 2.7~\textmu T in the center, measured with a fluxgate MAG-03 from Bartington Instruments. Before the current was switched on, we measured the remanent field along the $y$-axis, shown as the grey dotted line in Fig.~\ref{fig:Opti}(b). This measured remanent field with a gradient around 3 pT/cm was subtracted from the measured field when the current to the coils was on. Before each measurement, the BMSR-2 was degaussed in order to reduce the remanent field of the shield \cite{Thiel2007}, providing reproducible starting conditions after the change of the spacing distance between the coils. 

The measured residual-field-corrected curve of the coil in a 9 cm range is shown for three coil distances $d$ in Fig.~\ref{fig:Opti}(b). The curve with $d$ = 1.35 m (blue) corresponds to the expected optimum. To validate the expected strong dependency on the distance, the measurement was repeated for $d$ = 1.32 m and $d$ = 1.38 m. The mechanical accuracy in the alignment of the provisional setup of the coil leads to the observed shift of the field extremum away from $y=0$. This shift is different each time the coil distance is changed. To compare the measurements always a 5~cm regions starting from the extremum was chosen. The limited mechanical accuracy of the temporary setup also worsens also the homogeneity so that we expect to underestimate the possible improvement. Even for the final coil setup with a higher mechanical accuracy an adjustment of the experiment to the location of the most homogenous magnetic field is possible. For the blue curve, the most uniform 5~cm wide region is from -1.6 cm (where the field is strongest) to 3.4 cm with a field change of only 18 pT. This is a relative change of 6~ppm in relation to the maximal field strength of 2.7 \textmu T, which is 5.4 times smaller than that of the previously used circular Helmholtz coil. Comparing the results for the three distances, Fig.~\ref{fig:Opti}(b) confirms that the field homogeneity around the optimal value close to $d$ = 1.35~m deteriorates fast. A shift of 3~cm to 1.32~m or 1.38~m increases the field change in a 5~cm region starting from the extremum from 18~pT to 114~pT or 190~pT. The change from a maximum field strength for $d=1.38$ m to a local minimum field strength for $d=1.32$~m is a known behavior for such a coil arrangement. In our temporary setup it was difficult to move the coils and to align them in parallel, leading to an additional shift of the position of the field extremum as seen in Fig.~\ref{fig:Opti}(b). The gradient of the $d$ = 1.35 m curve is already close to the gradient of the residual field (remanence). It is unclear if the residual field contribution is unchanged on this level when the coil is turned on and the inner layer of the shield is degaussed. We subtracted the residual field even though the raw data show a more homogeneous field. 

By designing a larger and optimized coil set, we significantly improved the uniformity of the coil field, compared to the circular Helmholtz coil. Another advantage of the square coil prototype is that a noise peak at 4~Hz due to vibrations disappeared because the coils were attached to the walls. Note that the distance between the cables and the MSR walls is only a few~cm. Close to the cables the shielding material is exposed to strong fields. It even could reach saturation. Therefore, $\mu_\text{r}$ will decrease near the cables, leading to a reduction of the shielding factor. It was measured that when the coil pair is on, the low-frequency shielding factor of the BMSR-2 decreases by around 5\%. For the $^{129}\text{Xe}$-EDM experiment this is acceptable.

\section{Conclusions}

The proposed three-step model takes into account the impact of a finite-thickness and finite-permeability magnetic shield on the field generated by internal coils. To deal with the finite-thickness plates parallel to the coil, we developed an improved mirror method as an extension of the classical images method. The three-step model is significantly faster than a pure FEM analysis with still reasonable accuracy, which is important for optimizing coil spacings. 

The three-step model was verified with the FEM calculation for a square coil pair inside an MSR. In this case the relative error of the three-step model on the magnetic field strength is in the order of the expected 0.1\% accuracy of the FEM calculation used as benchmark. We applied the model to calculate the change of the magnetic field created by a 1.6-m-diameter circular Helmholtz coil inside BMSR-2 and obtained a result consistent with the measurement data. 

As an example of coil optimization, the optimal distance between two square coils for the best homogeneity in a defined region along the central axis was calculated. We built a coil prototype of 2.75~m in side length attached to the BMSR-2 walls. The optimal distance was rapidly estimated by the three-step model and verified by FEM calculation as well as by experiment. The experimental result of the prototype coil shows that the optimum distance could be predicted within the geometric uncertainty of the setup. The maximal change of the magnetic field is 18~pT over a 5~cm region near the center at the field strength of 2.7 \textmu T, corresponding to a relative field change of 6~ppm. This is more than a factor of 5 smaller than the value measured for the previously used Helmholtz coil. Due to this result we have already optimized a built-in coil set with four square coils with the three-step method. This coil system will be installed inside BMSR-2 during a planned upgrade with a new innermost shielding layer.

\section*{Acknowledgments}
This work was supported by the Funds for International Cooperation and Exchange of NSFC-DFG (Grant \linebreak No.51861135308). We acknowledge the support of the Core Facility ‘Metrology of Ultra-Low Magnetic Fields’ at \linebreak Physikalisch-Technische Bundesanstalt which receives funding from the Deutsche Forschungsgemeinschaft (DFG KO 5321/3-1 and TR 408/11-1). T. Liu acknowledges the support from Chinese Scholarship Commitment (Grant No. \linebreak 201706120197). Z. Sun acknowledges the Yong Scientists Fund of the National Natural Science Foundation of China (Grant No. 51807039) and the International Postdoctoral Exchange Fellowship Program (Grant No.20171023).

\bibliography{ms}

\providecommand{\noopsort}[1]{}\providecommand{\singleletter}[1]{#1}%
\begin{thebibliography}{42}%
\makeatletter
\providecommand \@ifxundefined [1]{%
 \@ifx{#1\undefined}
}%
\providecommand \@ifnum [1]{%
 \ifnum #1\expandafter \@firstoftwo
 \else \expandafter \@secondoftwo
 \fi
}%
\providecommand \@ifx [1]{%
 \ifx #1\expandafter \@firstoftwo
 \else \expandafter \@secondoftwo
 \fi
}%
\providecommand \natexlab [1]{#1}%
\providecommand \enquote  [1]{``#1''}%
\providecommand \bibnamefont  [1]{#1}%
\providecommand \bibfnamefont [1]{#1}%
\providecommand \citenamefont [1]{#1}%
\providecommand \href@noop [0]{\@secondoftwo}%
\providecommand \href [0]{\begingroup \@sanitize@url \@href}%
\providecommand \@href[1]{\@@startlink{#1}\@@href}%
\providecommand \@@href[1]{\endgroup#1\@@endlink}%
\providecommand \@sanitize@url [0]{\catcode `\\12\catcode `\$12\catcode
  `\&12\catcode `\#12\catcode `\^12\catcode `\_12\catcode `\%12\relax}%
\providecommand \@@startlink[1]{}%
\providecommand \@@endlink[0]{}%
\providecommand \url  [0]{\begingroup\@sanitize@url \@url }%
\providecommand \@url [1]{\endgroup\@href {#1}{\urlprefix }}%
\providecommand \urlprefix  [0]{URL }%
\providecommand \Eprint [0]{\href }%
\providecommand \doibase [0]{https://doi.org/}%
\providecommand \selectlanguage [0]{\@gobble}%
\providecommand \bibinfo  [0]{\@secondoftwo}%
\providecommand \bibfield  [0]{\@secondoftwo}%
\providecommand \translation [1]{[#1]}%
\providecommand \BibitemOpen [0]{}%
\providecommand \bibitemStop [0]{}%
\providecommand \bibitemNoStop [0]{.\EOS\space}%
\providecommand \EOS [0]{\spacefactor3000\relax}%
\providecommand \BibitemShut  [1]{\csname bibitem#1\endcsname}%
\let\auto@bib@innerbib\@empty
\bibitem [{\citenamefont {Sakamoto}\ \emph {et~al.}(2015)\citenamefont
  {Sakamoto} \emph {et~al.}}]{Sakamoto2015}%
  \BibitemOpen
  \bibfield  {author} {\bibinfo {author} {\bibfnamefont {Y.}~\bibnamefont
  {Sakamoto}} \emph {et~al.},\ }\bibfield  {title} {\enquote {\bibinfo {title}
  {{Development of high-homogeneity magnetic field coil for $^{129}\text{Xe}$
  EDM experiment}},}\ }\href {https://doi.org/10.1007/s10751-014-1109-5}
  {\bibfield  {journal} {\bibinfo  {journal} {Hyperfine Interactions}\ }\textbf
  {\bibinfo {volume} {230}},\ \bibinfo {pages} {141--146} (\bibinfo {year}
  {2015})}\BibitemShut {NoStop}%
\bibitem [{\citenamefont {Slutsky}\ \emph {et~al.}(2017)\citenamefont {Slutsky}
  \emph {et~al.}}]{Slutsky2017}%
  \BibitemOpen
  \bibfield  {author} {\bibinfo {author} {\bibfnamefont {S.}~\bibnamefont
  {Slutsky}} \emph {et~al.},\ }\bibfield  {title} {\enquote {\bibinfo {title}
  {{Cryogenic magnetic coil and superconducting magnetic shield for neutron
  electric dipole moment searches}},}\ }\href
  {https://doi.org/10.1016/j.nima.2017.05.005} {\bibfield  {journal} {\bibinfo
  {journal} {Nuclear Instruments and Methods in Physics Research, Section A:
  Accelerators, Spectrometers, Detectors and Associated Equipment}\ }\textbf
  {\bibinfo {volume} {862}},\ \bibinfo {pages} {36--48} (\bibinfo {year}
  {2017})}\BibitemShut {NoStop}%
\bibitem [{\citenamefont {{P{\'{e}}rez Galv{\'{a}}n}}\ \emph
  {et~al.}(2011)\citenamefont {{P{\'{e}}rez Galv{\'{a}}n}} \emph
  {et~al.}}]{PerezGalvan2011}%
  \BibitemOpen
  \bibfield  {author} {\bibinfo {author} {\bibfnamefont {A.}~\bibnamefont
  {{P{\'{e}}rez Galv{\'{a}}n}}} \emph {et~al.},\ }\bibfield  {title} {\enquote
  {\bibinfo {title} {{High uniformity magnetic coil for search of neutron
  electric dipole moment}},}\ }\href
  {https://doi.org/10.1016/j.nima.2011.09.019} {\bibfield  {journal} {\bibinfo
  {journal} {Nuclear Instruments and Methods in Physics Research, Section A:
  Accelerators, Spectrometers, Detectors and Associated Equipment}\ }\textbf
  {\bibinfo {volume} {660}},\ \bibinfo {pages} {147--153} (\bibinfo {year}
  {2011})}\BibitemShut {NoStop}%
\bibitem [{\citenamefont {Abe}\ \emph {et~al.}(2018)\citenamefont {Abe} \emph
  {et~al.}}]{Abe2018}%
  \BibitemOpen
  \bibfield  {author} {\bibinfo {author} {\bibfnamefont {M.}~\bibnamefont
  {Abe}} \emph {et~al.},\ }\bibfield  {title} {\enquote {\bibinfo {title}
  {{Magnetic design and method of a superconducting magnet for muon g-2/EDM
  precise measurements in a cylindrical volume with homogeneous magnetic
  field}},}\ }\href {https://doi.org/10.1016/j.nima.2018.01.026} {\bibfield
  {journal} {\bibinfo  {journal} {Nuclear Instruments and Methods in Physics
  Research, Section A: Accelerators, Spectrometers, Detectors and Associated
  Equipment}\ }\textbf {\bibinfo {volume} {890}},\ \bibinfo {pages} {51--63}
  (\bibinfo {year} {2018})}\BibitemShut {NoStop}%
\bibitem [{\citenamefont {Zikmund}\ \emph {et~al.}(2015)\citenamefont
  {Zikmund}, \citenamefont {Ripka}, \citenamefont {Ketzler}, \citenamefont
  {Harcken},\ and\ \citenamefont {Albrecht}}]{Zikmund2015a}%
  \BibitemOpen
  \bibfield  {author} {\bibinfo {author} {\bibfnamefont {A.}~\bibnamefont
  {Zikmund}}, \bibinfo {author} {\bibfnamefont {P.}~\bibnamefont {Ripka}},
  \bibinfo {author} {\bibfnamefont {R.}~\bibnamefont {Ketzler}}, \bibinfo
  {author} {\bibfnamefont {H.}~\bibnamefont {Harcken}},\ and\ \bibinfo {author}
  {\bibfnamefont {M.}~\bibnamefont {Albrecht}},\ }\bibfield  {title} {\enquote
  {\bibinfo {title} {{Precise scalar calibration of a tri-axial Braunbek coil
  system}},}\ }\href {https://doi.org/10.1109/TMAG.2014.2357783} {\bibfield
  {journal} {\bibinfo  {journal} {IEEE Transactions on Magnetics}\ }\textbf
  {\bibinfo {volume} {51}},\ \bibinfo {pages} {3--6} (\bibinfo {year}
  {2015})}\BibitemShut {NoStop}%
\bibitem [{\citenamefont {Bronaugh}(1995)}]{Bronaugh1995}%
  \BibitemOpen
  \bibfield  {author} {\bibinfo {author} {\bibfnamefont {E.}~\bibnamefont
  {Bronaugh}},\ }\bibfield  {title} {\enquote {\bibinfo {title} {{Helmholtz
  coils for calibration of probes and sensors: limits of magnetic field
  accuracy and uniformity}},}\ }in\ \href
  {https://doi.org/10.1109/ISEMC.1995.523521} {\emph {\bibinfo {booktitle}
  {Proceedings of International Symposium on Electromagnetic Compatibility}}}\
  (\bibinfo  {publisher} {IEEE},\ \bibinfo {year} {1995})\ pp.\ \bibinfo
  {pages} {72--76}\BibitemShut {NoStop}%
\bibitem [{\citenamefont {Wang}\ \emph {et~al.}(2019)\citenamefont {Wang},
  \citenamefont {Zhou}, \citenamefont {Liu}, \citenamefont {Wu}, \citenamefont
  {Chen}, \citenamefont {Han},\ and\ \citenamefont {Fang}}]{Wang2019}%
  \BibitemOpen
  \bibfield  {author} {\bibinfo {author} {\bibfnamefont {J.}~\bibnamefont
  {Wang}}, \bibinfo {author} {\bibfnamefont {B.}~\bibnamefont {Zhou}}, \bibinfo
  {author} {\bibfnamefont {X.}~\bibnamefont {Liu}}, \bibinfo {author}
  {\bibfnamefont {W.}~\bibnamefont {Wu}}, \bibinfo {author} {\bibfnamefont
  {L.}~\bibnamefont {Chen}}, \bibinfo {author} {\bibfnamefont {B.}~\bibnamefont
  {Han}},\ and\ \bibinfo {author} {\bibfnamefont {J.}~\bibnamefont {Fang}},\
  }\bibfield  {title} {\enquote {\bibinfo {title} {{An Improved Target-Field
  Method for the Design of Uniform Magnetic Field Coils in Miniature Atomic
  Sensors}},}\ }\href {https://doi.org/10.1109/ACCESS.2019.2920955} {\bibfield
  {journal} {\bibinfo  {journal} {IEEE Access}\ }\textbf {\bibinfo {volume}
  {7}},\ \bibinfo {pages} {74800--74810} (\bibinfo {year} {2019})}\BibitemShut
  {NoStop}%
\bibitem [{\citenamefont {Voigt}\ \emph {et~al.}(2013)\citenamefont {Voigt},
  \citenamefont {Knappe-Gr{\"{u}}neberg}, \citenamefont {Schnabel},
  \citenamefont {K{\"{o}}rber},\ and\ \citenamefont {Burghoff}}]{Voigt2013}%
  \BibitemOpen
  \bibfield  {author} {\bibinfo {author} {\bibfnamefont {J.}~\bibnamefont
  {Voigt}}, \bibinfo {author} {\bibfnamefont {S.}~\bibnamefont
  {Knappe-Gr{\"{u}}neberg}}, \bibinfo {author} {\bibfnamefont {A.}~\bibnamefont
  {Schnabel}}, \bibinfo {author} {\bibfnamefont {R.}~\bibnamefont
  {K{\"{o}}rber}},\ and\ \bibinfo {author} {\bibfnamefont {M.}~\bibnamefont
  {Burghoff}},\ }\bibfield  {title} {\enquote {\bibinfo {title} {{Measures to
  reduce the residual field and field gradient inside a magnetically shielded
  room by a factor of more than 10}},}\ }\href
  {https://doi.org/10.2478/mms-2013-0021} {\bibfield  {journal} {\bibinfo
  {journal} {Metrology and Measurement Systems}\ }\textbf {\bibinfo {volume}
  {20}},\ \bibinfo {pages} {239--248} (\bibinfo {year} {2013})}\BibitemShut
  {NoStop}%
\bibitem [{\citenamefont {Yamazaki}\ \emph {et~al.}(2009)\citenamefont
  {Yamazaki}, \citenamefont {Abe}, \citenamefont {Terazono}, \citenamefont
  {Fujimaki}, \citenamefont {Murata}, \citenamefont {Oyama},\ and\
  \citenamefont {Kobayashi}}]{Yamazaki2009}%
  \BibitemOpen
  \bibfield  {author} {\bibinfo {author} {\bibfnamefont {K.}~\bibnamefont
  {Yamazaki}}, \bibinfo {author} {\bibfnamefont {T.}~\bibnamefont {Abe}},
  \bibinfo {author} {\bibfnamefont {Y.}~\bibnamefont {Terazono}}, \bibinfo
  {author} {\bibfnamefont {N.}~\bibnamefont {Fujimaki}}, \bibinfo {author}
  {\bibfnamefont {T.}~\bibnamefont {Murata}}, \bibinfo {author} {\bibfnamefont
  {D.}~\bibnamefont {Oyama}},\ and\ \bibinfo {author} {\bibfnamefont
  {K.}~\bibnamefont {Kobayashi}},\ }\bibfield  {title} {\enquote {\bibinfo
  {title} {{Magnetic Noise Due to Sound of Footsteps on Wooden Free-Access
  Floor Outside a Magnetically Shielded Room for Biomagnetic and Nondestructive
  Measurements}},}\ }\href {https://doi.org/10.1109/TMAG.2009.2025185}
  {\bibfield  {journal} {\bibinfo  {journal} {IEEE Transactions on Magnetics}\
  }\textbf {\bibinfo {volume} {45}},\ \bibinfo {pages} {4644--4647} (\bibinfo
  {year} {2009})}\BibitemShut {NoStop}%
\bibitem [{\citenamefont {Clayton}(2011)}]{Clayton2011}%
  \BibitemOpen
  \bibfield  {author} {\bibinfo {author} {\bibfnamefont {S.~M.}\ \bibnamefont
  {Clayton}},\ }\bibfield  {title} {\enquote {\bibinfo {title} {{Spin
  relaxation and linear-in-electric-field frequency shift in an arbitrary,
  time-independent magnetic field}},}\ }\href
  {https://doi.org/10.1016/j.jmr.2011.04.008} {\bibfield  {journal} {\bibinfo
  {journal} {Journal of Magnetic Resonance}\ }\textbf {\bibinfo {volume}
  {211}},\ \bibinfo {pages} {89--95} (\bibinfo {year} {2011})}\BibitemShut
  {NoStop}%
\bibitem [{\citenamefont {Cates}, \citenamefont {Schaefer},\ and\ \citenamefont
  {Happer}(1988)}]{Cates1988}%
  \BibitemOpen
  \bibfield  {author} {\bibinfo {author} {\bibfnamefont {G.~D.}\ \bibnamefont
  {Cates}}, \bibinfo {author} {\bibfnamefont {S.~R.}\ \bibnamefont
  {Schaefer}},\ and\ \bibinfo {author} {\bibfnamefont {W.}~\bibnamefont
  {Happer}},\ }\bibfield  {title} {\enquote {\bibinfo {title} {{Relaxation of
  spins due to field inhomogeneities in gaseous samples at low magnetic fields
  and low pressures}},}\ }\href {https://doi.org/10.1103/PhysRevA.37.2877}
  {\bibfield  {journal} {\bibinfo  {journal} {Physical Review A}\ }\textbf
  {\bibinfo {volume} {37}},\ \bibinfo {pages} {2877--2885} (\bibinfo {year}
  {1988})}\BibitemShut {NoStop}%
\bibitem [{\citenamefont {Allmendinger}\ \emph {et~al.}(2017)\citenamefont
  {Allmendinger} \emph {et~al.}}]{Allmendinger2017}%
  \BibitemOpen
  \bibfield  {author} {\bibinfo {author} {\bibfnamefont {F.}~\bibnamefont
  {Allmendinger}} \emph {et~al.},\ }\bibfield  {title} {\enquote {\bibinfo
  {title} {{Precise measurement of magnetic field gradients from free spin
  precession signals of 3He and 129Xe magnetometers}},}\ }\href
  {https://doi.org/10.1140/epjd/e2017-70505-4} {\bibfield  {journal} {\bibinfo
  {journal} {The European Physical Journal D}\ }\textbf {\bibinfo {volume}
  {71}},\ \bibinfo {pages} {98} (\bibinfo {year} {2017})}\BibitemShut {NoStop}%
\bibitem [{\citenamefont {Afach}\ \emph {et~al.}(2015)\citenamefont {Afach}
  \emph {et~al.}}]{Afach2015}%
  \BibitemOpen
  \bibfield  {author} {\bibinfo {author} {\bibfnamefont {S.}~\bibnamefont
  {Afach}} \emph {et~al.},\ }\bibfield  {title} {\enquote {\bibinfo {title}
  {{Measurement of a false electric dipole moment signal from 199Hg atoms
  exposed to an inhomogeneous magnetic field}},}\ }\href
  {https://doi.org/10.1140/epjd/e2015-60207-4} {\bibfield  {journal} {\bibinfo
  {journal} {European Physical Journal D}\ }\textbf {\bibinfo {volume} {69}},\
  \bibinfo {pages} {225} (\bibinfo {year} {2015})}\BibitemShut {NoStop}%
\bibitem [{\citenamefont {Pignol}\ and\ \citenamefont
  {Roccia}(2012)}]{Pignol2012}%
  \BibitemOpen
  \bibfield  {author} {\bibinfo {author} {\bibfnamefont {G.}~\bibnamefont
  {Pignol}}\ and\ \bibinfo {author} {\bibfnamefont {S.}~\bibnamefont
  {Roccia}},\ }\bibfield  {title} {\enquote {\bibinfo {title}
  {{Electric-dipole-moment searches: Reexamination of frequency shifts for
  particles in traps}},}\ }\href {https://doi.org/10.1103/PhysRevA.85.042105}
  {\bibfield  {journal} {\bibinfo  {journal} {Physical Review A}\ }\textbf
  {\bibinfo {volume} {85}},\ \bibinfo {pages} {042105} (\bibinfo {year}
  {2012})}\BibitemShut {NoStop}%
\bibitem [{\citenamefont {Abel}\ \emph {et~al.}(2020)\citenamefont {Abel} \emph
  {et~al.}}]{Abel2020}%
  \BibitemOpen
  \bibfield  {author} {\bibinfo {author} {\bibfnamefont {C.}~\bibnamefont
  {Abel}} \emph {et~al.},\ }\bibfield  {title} {\enquote {\bibinfo {title}
  {{Measurement of the Permanent Electric Dipole Moment of the Neutron}},}\
  }\href {https://doi.org/10.1103/PhysRevLett.124.081803} {\bibfield  {journal}
  {\bibinfo  {journal} {Physical Review Letters}\ }\textbf {\bibinfo {volume}
  {124}},\ \bibinfo {pages} {081803} (\bibinfo {year} {2020})}\BibitemShut
  {NoStop}%
\bibitem [{\citenamefont {Abel}\ \emph {et~al.}(2019)\citenamefont {Abel} \emph
  {et~al.}}]{Abel2019a}%
  \BibitemOpen
  \bibfield  {author} {\bibinfo {author} {\bibfnamefont {C.}~\bibnamefont
  {Abel}} \emph {et~al.},\ }\bibfield  {title} {\enquote {\bibinfo {title}
  {{Magnetic-field uniformity in neutron electric-dipole-moment
  experiments}},}\ }\href {https://doi.org/10.1103/PhysRevA.99.042112}
  {\bibfield  {journal} {\bibinfo  {journal} {Physical Review A}\ }\textbf
  {\bibinfo {volume} {99}},\ \bibinfo {pages} {1--16} (\bibinfo {year}
  {2019})}\BibitemShut {NoStop}%
\bibitem [{\citenamefont {Dadisman}(2018)}]{Dadisman2018}%
  \BibitemOpen
  \bibfield  {author} {\bibinfo {author} {\bibfnamefont {J.~R.}\ \bibnamefont
  {Dadisman}},\ }\emph {\bibinfo {title} {{Magnetic field design to reduce
  systematic effects in neutron electric dipole moment measurements}}},\ \href
  {https://doi.org/doi.org/10.13023/ETD.2018.094} {\bibinfo {type} {Theses and
  dissertations--physics and astronomy}},\ \bibinfo  {school} {University of
  Kentucky} (\bibinfo {year} {2018})\BibitemShut {NoStop}%
\bibitem [{\citenamefont {Hosoya}\ and\ \citenamefont
  {Goto}(1991)}]{Hosoya1991}%
  \BibitemOpen
  \bibfield  {author} {\bibinfo {author} {\bibfnamefont {M.}~\bibnamefont
  {Hosoya}}\ and\ \bibinfo {author} {\bibfnamefont {E.}~\bibnamefont {Goto}},\
  }\bibfield  {title} {\enquote {\bibinfo {title} {{Coils for generating
  uniform fields in a cylindrical ferromagnetic shield}},}\ }\href
  {https://doi.org/10.1063/1.1142267} {\bibfield  {journal} {\bibinfo
  {journal} {Review of Scientific Instruments}\ }\textbf {\bibinfo {volume}
  {62}},\ \bibinfo {pages} {2472--2475} (\bibinfo {year} {1991})}\BibitemShut
  {NoStop}%
\bibitem [{\citenamefont {Wyszy{\'{n}}ski}\ \emph {et~al.}(2017)\citenamefont
  {Wyszy{\'{n}}ski} \emph {et~al.}}]{Wyszynski2017}%
  \BibitemOpen
  \bibfield  {author} {\bibinfo {author} {\bibfnamefont {G.}~\bibnamefont
  {Wyszy{\'{n}}ski}} \emph {et~al.},\ }\bibfield  {title} {\enquote {\bibinfo
  {title} {{Active compensation of magnetic field distortions based on vector
  spherical harmonics field description}},}\ }\href
  {https://doi.org/10.1063/1.4978394} {\bibfield  {journal} {\bibinfo
  {journal} {AIP Advances}\ }\textbf {\bibinfo {volume} {7}},\ \bibinfo {pages}
  {035216} (\bibinfo {year} {2017})}\BibitemShut {NoStop}%
\bibitem [{\citenamefont {Altarev}\ \emph {et~al.}(2014)\citenamefont {Altarev}
  \emph {et~al.}}]{Altarev2014}%
  \BibitemOpen
  \bibfield  {author} {\bibinfo {author} {\bibfnamefont {I.}~\bibnamefont
  {Altarev}} \emph {et~al.},\ }\bibfield  {title} {\enquote {\bibinfo {title}
  {{A magnetically shielded room with ultra low residual field and
  gradient}},}\ }\href {https://doi.org/10.1063/1.4886146} {\bibfield
  {journal} {\bibinfo  {journal} {Review of Scientific Instruments}\ }\textbf
  {\bibinfo {volume} {85}},\ \bibinfo {pages} {075106} (\bibinfo {year}
  {2014})}\BibitemShut {NoStop}%
\bibitem [{\citenamefont {Liu}\ \emph {et~al.}(2020{\natexlab{a}})\citenamefont
  {Liu}, \citenamefont {Andalib}, \citenamefont {Ostapchuk},\ and\
  \citenamefont {Bidinosti}}]{Liu2020}%
  \BibitemOpen
  \bibfield  {author} {\bibinfo {author} {\bibfnamefont {C.-Y.}\ \bibnamefont
  {Liu}}, \bibinfo {author} {\bibfnamefont {T.}~\bibnamefont {Andalib}},
  \bibinfo {author} {\bibfnamefont {D.}~\bibnamefont {Ostapchuk}},\ and\
  \bibinfo {author} {\bibfnamefont {C.}~\bibnamefont {Bidinosti}},\ }\bibfield
  {title} {\enquote {\bibinfo {title} {{Analytic models of magnetically
  enclosed spherical and solenoidal coils}},}\ }\href
  {https://doi.org/10.1016/j.nima.2019.162837} {\bibfield  {journal} {\bibinfo
  {journal} {Nuclear Instruments and Methods in Physics Research Section A:
  Accelerators, Spectrometers, Detectors and Associated Equipment}\ }\textbf
  {\bibinfo {volume} {949}},\ \bibinfo {pages} {162837} (\bibinfo {year}
  {2020}{\natexlab{a}})}\BibitemShut {NoStop}%
\bibitem [{\citenamefont {Nouri}\ and\ \citenamefont
  {Plaster}(2013)}]{Nouri2013}%
  \BibitemOpen
  \bibfield  {author} {\bibinfo {author} {\bibfnamefont {N.}~\bibnamefont
  {Nouri}}\ and\ \bibinfo {author} {\bibfnamefont {B.}~\bibnamefont
  {Plaster}},\ }\bibfield  {title} {\enquote {\bibinfo {title} {{Comparison of
  magnetic field uniformities for discretized and finite-sized standard cos
  $\theta$, solenoidal, and spherical coils}},}\ }\href
  {https://doi.org/10.1016/j.nima.2013.05.013} {\bibfield  {journal} {\bibinfo
  {journal} {Nuclear Instruments and Methods in Physics Research, Section A:
  Accelerators, Spectrometers, Detectors and Associated Equipment}\ }\textbf
  {\bibinfo {volume} {723}},\ \bibinfo {pages} {30--35} (\bibinfo {year}
  {2013})}\BibitemShut {NoStop}%
\bibitem [{\citenamefont {Wu}\ \emph {et~al.}(2019)\citenamefont {Wu},
  \citenamefont {Zhou}, \citenamefont {Liu}, \citenamefont {Wang},
  \citenamefont {Pang}, \citenamefont {Chen}, \citenamefont {Quan},\ and\
  \citenamefont {Liu}}]{Wu2019}%
  \BibitemOpen
  \bibfield  {author} {\bibinfo {author} {\bibfnamefont {W.}~\bibnamefont
  {Wu}}, \bibinfo {author} {\bibfnamefont {B.}~\bibnamefont {Zhou}}, \bibinfo
  {author} {\bibfnamefont {Z.}~\bibnamefont {Liu}}, \bibinfo {author}
  {\bibfnamefont {J.}~\bibnamefont {Wang}}, \bibinfo {author} {\bibfnamefont
  {H.}~\bibnamefont {Pang}}, \bibinfo {author} {\bibfnamefont {L.}~\bibnamefont
  {Chen}}, \bibinfo {author} {\bibfnamefont {W.}~\bibnamefont {Quan}},\ and\
  \bibinfo {author} {\bibfnamefont {G.}~\bibnamefont {Liu}},\ }\bibfield
  {title} {\enquote {\bibinfo {title} {{Design of Highly Uniform Magnetic Field
  Coils Based on a Particle Swarm Optimization Algorithm}},}\ }\href
  {https://doi.org/10.1109/access.2019.2933608} {\bibfield  {journal} {\bibinfo
   {journal} {IEEE Access}\ }\textbf {\bibinfo {volume} {7}},\ \bibinfo {pages}
  {125310--125322} (\bibinfo {year} {2019})}\BibitemShut {NoStop}%
\bibitem [{\citenamefont {Hanson}\ and\ \citenamefont
  {Pipkin}(1965)}]{Hanson1965}%
  \BibitemOpen
  \bibfield  {author} {\bibinfo {author} {\bibfnamefont {R.~J.}\ \bibnamefont
  {Hanson}}\ and\ \bibinfo {author} {\bibfnamefont {F.~M.}\ \bibnamefont
  {Pipkin}},\ }\bibfield  {title} {\enquote {\bibinfo {title} {{Magnetically
  shielded solenoid with field of high homogeneity}},}\ }\href
  {https://doi.org/10.1063/1.1719514} {\bibfield  {journal} {\bibinfo
  {journal} {Review of Scientific Instruments}\ }\textbf {\bibinfo {volume}
  {36}},\ \bibinfo {pages} {179--188} (\bibinfo {year} {1965})}\BibitemShut
  {NoStop}%
\bibitem [{\citenamefont {Bidinosti}, \citenamefont {Sakamoto},\ and\
  \citenamefont {Asahi}(2014)}]{Bidinosti2014a}%
  \BibitemOpen
  \bibfield  {author} {\bibinfo {author} {\bibfnamefont {C.~P.}\ \bibnamefont
  {Bidinosti}}, \bibinfo {author} {\bibfnamefont {Y.}~\bibnamefont
  {Sakamoto}},\ and\ \bibinfo {author} {\bibfnamefont {K.}~\bibnamefont
  {Asahi}},\ }\bibfield  {title} {\enquote {\bibinfo {title} {{General Solution
  of the Hollow Cylinder and Concentric DC Surface Current}},}\ }\href
  {https://doi.org/10.1109/LMAG.2014.2330346} {\bibfield  {journal} {\bibinfo
  {journal} {IEEE Magnetics Letters}\ }\textbf {\bibinfo {volume} {5}},\
  \bibinfo {pages} {1--4} (\bibinfo {year} {2014})}\BibitemShut {NoStop}%
\bibitem [{\citenamefont {Liu}\ \emph {et~al.}(2020{\natexlab{b}})\citenamefont
  {Liu}, \citenamefont {Schnabel}, \citenamefont {Sun}, \citenamefont {Voigt},\
  and\ \citenamefont {Li}}]{Liu2020a}%
  \BibitemOpen
  \bibfield  {author} {\bibinfo {author} {\bibfnamefont {T.}~\bibnamefont
  {Liu}}, \bibinfo {author} {\bibfnamefont {A.}~\bibnamefont {Schnabel}},
  \bibinfo {author} {\bibfnamefont {Z.}~\bibnamefont {Sun}}, \bibinfo {author}
  {\bibfnamefont {J.}~\bibnamefont {Voigt}},\ and\ \bibinfo {author}
  {\bibfnamefont {L.}~\bibnamefont {Li}},\ }\bibfield  {title} {\enquote
  {\bibinfo {title} {{Approximate expressions for the magnetic field created by
  circular coils inside a closed cylindrical shield of finite thickness and
  permeability}},}\ }\href {https://doi.org/10.1016/j.jmmm.2020.166846}
  {\bibfield  {journal} {\bibinfo  {journal} {Journal of Magnetism and Magnetic
  Materials}\ }\textbf {\bibinfo {volume} {507}},\ \bibinfo {pages} {166846}
  (\bibinfo {year} {2020}{\natexlab{b}})}\BibitemShut {NoStop}%
\bibitem [{\citenamefont {Turouski}\ and\ \citenamefont
  {Turouski}(2013)}]{inbookTurouski2013}%
  \BibitemOpen
  \bibfield  {author} {\bibinfo {author} {\bibfnamefont {J.}~\bibnamefont
  {Turouski}}\ and\ \bibinfo {author} {\bibfnamefont {M.}~\bibnamefont
  {Turouski}},\ }\enquote {\bibinfo {title} {{Electric Machine, Transformer,
  and Power Equipment Design}},}\ \ (\bibinfo  {publisher} {CRC press},\
  \bibinfo {year} {2013})\ pp.\ \bibinfo {pages} {243--258}\BibitemShut
  {NoStop}%
\bibitem [{\citenamefont {Liu}\ \emph {et~al.}(2018)\citenamefont {Liu},
  \citenamefont {Voigt}, \citenamefont {Sun}, \citenamefont {Schnabel},
  \citenamefont {Knappe-Grueneberg}, \citenamefont {Fan},\ and\ \citenamefont
  {Li}}]{Liu2018}%
  \BibitemOpen
  \bibfield  {author} {\bibinfo {author} {\bibfnamefont {T.}~\bibnamefont
  {Liu}}, \bibinfo {author} {\bibfnamefont {J.}~\bibnamefont {Voigt}}, \bibinfo
  {author} {\bibfnamefont {Z.}~\bibnamefont {Sun}}, \bibinfo {author}
  {\bibfnamefont {A.}~\bibnamefont {Schnabel}}, \bibinfo {author} {\bibnamefont
  {Knappe-Grueneberg}}, \bibinfo {author} {\bibfnamefont {I.}~\bibnamefont
  {Fan}},\ and\ \bibinfo {author} {\bibfnamefont {L.}~\bibnamefont {Li}},\
  }\bibfield  {title} {\enquote {\bibinfo {title} {{Two-Step Mirror Model for
  the Optimization of the Magnetic Field Generated by Coils Inside Magnetic
  Shielding}},}\ }in\ \href {https://doi.org/10.1109/CPEM.2018.8500866} {\emph
  {\bibinfo {booktitle} {2018 Conference on Precision Electromagnetic
  Measurements (CPEM 2018)}}}\ (\bibinfo  {publisher} {IEEE},\ \bibinfo {year}
  {2018})\ pp.\ \bibinfo {pages} {1--2}\BibitemShut {NoStop}%
\bibitem [{\citenamefont {Pan}\ \emph {et~al.}(2020)\citenamefont {Pan},
  \citenamefont {Lin}, \citenamefont {Li}, \citenamefont {Li}, \citenamefont
  {Jin}, \citenamefont {Sun},\ and\ \citenamefont {Liu}}]{Pan2019}%
  \BibitemOpen
  \bibfield  {author} {\bibinfo {author} {\bibfnamefont {D.}~\bibnamefont
  {Pan}}, \bibinfo {author} {\bibfnamefont {S.}~\bibnamefont {Lin}}, \bibinfo
  {author} {\bibfnamefont {L.}~\bibnamefont {Li}}, \bibinfo {author}
  {\bibfnamefont {J.}~\bibnamefont {Li}}, \bibinfo {author} {\bibfnamefont
  {Y.}~\bibnamefont {Jin}}, \bibinfo {author} {\bibfnamefont {Z.}~\bibnamefont
  {Sun}},\ and\ \bibinfo {author} {\bibfnamefont {T.}~\bibnamefont {Liu}},\
  }\bibfield  {title} {\enquote {\bibinfo {title} {{Research on the Design
  Method of Uniform Magnetic Field Coil Based on the MSR}},}\ }\href
  {https://doi.org/10.1109/TIE.2019.2899544} {\bibfield  {journal} {\bibinfo
  {journal} {IEEE Transactions on Industrial Electronics}\ }\textbf {\bibinfo
  {volume} {67}},\ \bibinfo {pages} {1348--1356} (\bibinfo {year}
  {2020})}\BibitemShut {NoStop}%
\bibitem [{\citenamefont {Lee}\ \emph {et~al.}(2013)\citenamefont {Lee},
  \citenamefont {Huh}, \citenamefont {Choi}, \citenamefont {Thai},
  \citenamefont {Kim}, \citenamefont {Al-Ammar}, \citenamefont {El-Kady},\ and\
  \citenamefont {Rim}}]{Lee2013}%
  \BibitemOpen
  \bibfield  {author} {\bibinfo {author} {\bibfnamefont {W.~Y.}\ \bibnamefont
  {Lee}}, \bibinfo {author} {\bibfnamefont {J.}~\bibnamefont {Huh}}, \bibinfo
  {author} {\bibfnamefont {S.~Y.}\ \bibnamefont {Choi}}, \bibinfo {author}
  {\bibfnamefont {X.~V.}\ \bibnamefont {Thai}}, \bibinfo {author}
  {\bibfnamefont {J.~H.}\ \bibnamefont {Kim}}, \bibinfo {author} {\bibfnamefont
  {E.~A.}\ \bibnamefont {Al-Ammar}}, \bibinfo {author} {\bibfnamefont {M.~A.}\
  \bibnamefont {El-Kady}},\ and\ \bibinfo {author} {\bibfnamefont {C.~T.}\
  \bibnamefont {Rim}},\ }\bibfield  {title} {\enquote {\bibinfo {title}
  {{Finite-width magnetic mirror models of mono and dual coils for wireless
  electric vehicles}},}\ }\href {https://doi.org/10.1109/TPEL.2012.2206404}
  {\bibfield  {journal} {\bibinfo  {journal} {IEEE Transactions on Power
  Electronics}\ }\textbf {\bibinfo {volume} {28}},\ \bibinfo {pages}
  {1413--1428} (\bibinfo {year} {2013})}\BibitemShut {NoStop}%
\bibitem [{\citenamefont {Bork}\ \emph {et~al.}(2001)\citenamefont {Bork},
  \citenamefont {Hahlbohm}, \citenamefont {Klein},\ and\ \citenamefont
  {Schnabel}}]{Bork2001}%
  \BibitemOpen
  \bibfield  {author} {\bibinfo {author} {\bibfnamefont {J.}~\bibnamefont
  {Bork}}, \bibinfo {author} {\bibfnamefont {H.}~\bibnamefont {Hahlbohm}},
  \bibinfo {author} {\bibfnamefont {R.}~\bibnamefont {Klein}},\ and\ \bibinfo
  {author} {\bibfnamefont {A.}~\bibnamefont {Schnabel}},\ }\bibfield  {title}
  {\enquote {\bibinfo {title} {{The 8-layered magnetically shielded room of the
  PTB : Design and construction}},}\ }in\ \href@noop {} {\emph {\bibinfo
  {booktitle} {Proceedings of the 12th international conference on
  Biomagnetism}}}\ (\bibinfo {year} {2001})\ pp.\ \bibinfo {pages}
  {970--973}\BibitemShut {NoStop}%
\bibitem [{\citenamefont {Abbott}(2015)}]{Abbott2015}%
  \BibitemOpen
  \bibfield  {author} {\bibinfo {author} {\bibfnamefont {J.~J.}\ \bibnamefont
  {Abbott}},\ }\bibfield  {title} {\enquote {\bibinfo {title} {{Parametric
  design of tri-axial nested Helmholtz coils}},}\ }\href
  {https://doi.org/10.1063/1.4919400} {\bibfield  {journal} {\bibinfo
  {journal} {Review of Scientific Instruments}\ }\textbf {\bibinfo {volume}
  {86}},\ \bibinfo {pages} {054701} (\bibinfo {year} {2015})}\BibitemShut
  {NoStop}%
\bibitem [{\citenamefont {Beiranvand}(2013)}]{Beiranvand2013}%
  \BibitemOpen
  \bibfield  {author} {\bibinfo {author} {\bibfnamefont {R.}~\bibnamefont
  {Beiranvand}},\ }\bibfield  {title} {\enquote {\bibinfo {title} {{Analyzing
  the uniformity of the generated magnetic field by a practical one-dimensional
  Helmholtz coils system}},}\ }\href {https://doi.org/10.1063/1.4813275}
  {\bibfield  {journal} {\bibinfo  {journal} {Review of Scientific
  Instruments}\ }\textbf {\bibinfo {volume} {84}},\ \bibinfo {pages} {075109}
  (\bibinfo {year} {2013})}\BibitemShut {NoStop}%
\bibitem [{\citenamefont {Celozzi}, \citenamefont {Araneo},\ and\ \citenamefont
  {Lovat}(2008)}]{inbookCelozzi2008}%
  \BibitemOpen
  \bibfield  {author} {\bibinfo {author} {\bibfnamefont {S.}~\bibnamefont
  {Celozzi}}, \bibinfo {author} {\bibfnamefont {R.}~\bibnamefont {Araneo}},\
  and\ \bibinfo {author} {\bibfnamefont {G.}~\bibnamefont {Lovat}},\ }\enquote
  {\bibinfo {title} {{Electromagnetic Shielding}},}\ in\ \href@noop {} {\emph
  {\bibinfo {booktitle} {Electromagnetic Compatibility}}},\ \bibinfo {editor}
  {edited by\ \bibinfo {editor} {\bibfnamefont {K.}~\bibnamefont {CHANG}}}\
  (\bibinfo  {publisher} {John Wiley {\&} Sons, Inc.},\ \bibinfo {address}
  {Hoboken},\ \bibinfo {year} {2008})\ pp.\ \bibinfo {pages}
  {301--305}\BibitemShut {NoStop}%
\bibitem [{\citenamefont {Jackson}(1998)}]{inbookJackson1998}%
  \BibitemOpen
  \bibfield  {author} {\bibinfo {author} {\bibfnamefont {J.~D.}\ \bibnamefont
  {Jackson}},\ }\enquote {\bibinfo {title} {{Classical electrodynamics}},}\ \
  (\bibinfo  {publisher} {Wiley},\ \bibinfo {address} {New York, NY},\ \bibinfo
  {year} {1998})\ pp.\ \bibinfo {pages} {182--183},\ \bibinfo {edition} {3rd}\
  ed.\BibitemShut {Stop}%
\bibitem [{\citenamefont {Hanson}\ and\ \citenamefont
  {Hirshman}(2002)}]{Hanson2002}%
  \BibitemOpen
  \bibfield  {author} {\bibinfo {author} {\bibfnamefont {J.~D.}\ \bibnamefont
  {Hanson}}\ and\ \bibinfo {author} {\bibfnamefont {S.~P.}\ \bibnamefont
  {Hirshman}},\ }\bibfield  {title} {\enquote {\bibinfo {title} {{Compact
  expressions for the Biot–Savart fields of a filamentary segment}},}\ }\href
  {https://doi.org/10.1063/1.1507589} {\bibfield  {journal} {\bibinfo
  {journal} {Physics of Plasmas}\ }\textbf {\bibinfo {volume} {9}},\ \bibinfo
  {pages} {4410--4412} (\bibinfo {year} {2002})}\BibitemShut {NoStop}%
\bibitem [{\citenamefont {{Walter Frei}}()}]{WalterFrei2014}%
  \BibitemOpen
  \bibfield  {author} {\bibinfo {author} {\bibnamefont {{Walter Frei}}},\
  }\href@noop {} {\enquote {\bibinfo {title} {{Exploiting Symmetry to Simplify
  Magnetic Field Modeling, available at
  https://www.comsol.com/blogs/exploiting-symmetry-simplify-magnetic-field-modeling/
  (accessed on March 2nd, 2020)}},}\ }\BibitemShut {NoStop}%
\bibitem [{\citenamefont {Firester}(1966)}]{Firester1966}%
  \BibitemOpen
  \bibfield  {author} {\bibinfo {author} {\bibfnamefont {A.~H.}\ \bibnamefont
  {Firester}},\ }\bibfield  {title} {\enquote {\bibinfo {title} {{Design of
  Square Helmholtz Coil Systems}},}\ }\href {https://doi.org/10.1063/1.1720478}
  {\bibfield  {journal} {\bibinfo  {journal} {Review of Scientific
  Instruments}\ }\textbf {\bibinfo {volume} {37}},\ \bibinfo {pages}
  {1264--1265} (\bibinfo {year} {1966})}\BibitemShut {NoStop}%
\bibitem [{\citenamefont {{Magnicon GmbH}}()}]{Magnicon_CSE}%
  \BibitemOpen
  \bibfield  {author} {\bibinfo {author} {\bibnamefont {{Magnicon GmbH}}},\
  }\href@noop {} {\enquote {\bibinfo {title} {{CSE-1 low-noise curent source,
  available at http://www.magnicon.com/squid-electronics/accessories/cse-1/}
  (accessed on january 2nd, 2020)},}\ }\BibitemShut {NoStop}%
\bibitem [{\citenamefont {Andalib}\ \emph {et~al.}(2017)\citenamefont
  {Andalib}, \citenamefont {Martin}, \citenamefont {Bidinosti}, \citenamefont
  {Mammei}, \citenamefont {Jamieson}, \citenamefont {Lang},\ and\ \citenamefont
  {Kikawa}}]{Andalib2017}%
  \BibitemOpen
  \bibfield  {author} {\bibinfo {author} {\bibfnamefont {T.}~\bibnamefont
  {Andalib}}, \bibinfo {author} {\bibfnamefont {J.}~\bibnamefont {Martin}},
  \bibinfo {author} {\bibfnamefont {C.}~\bibnamefont {Bidinosti}}, \bibinfo
  {author} {\bibfnamefont {R.}~\bibnamefont {Mammei}}, \bibinfo {author}
  {\bibfnamefont {B.}~\bibnamefont {Jamieson}}, \bibinfo {author}
  {\bibfnamefont {M.}~\bibnamefont {Lang}},\ and\ \bibinfo {author}
  {\bibfnamefont {T.}~\bibnamefont {Kikawa}},\ }\bibfield  {title} {\enquote
  {\bibinfo {title} {{Sensitivity of fields generated within magnetically
  shielded volumes to changes in magnetic permeability}},}\ }\href
  {https://doi.org/10.1016/j.nima.2017.05.050} {\bibfield  {journal} {\bibinfo
  {journal} {Nuclear Instruments and Methods in Physics Research Section A:
  Accelerators, Spectrometers, Detectors and Associated Equipment}\ }\textbf
  {\bibinfo {volume} {867}},\ \bibinfo {pages} {139--147} (\bibinfo {year}
  {2017})}\BibitemShut {NoStop}%
\bibitem [{\citenamefont {Sachdeva}\ \emph {et~al.}(2019)\citenamefont
  {Sachdeva} \emph {et~al.}}]{Sachdeva2019}%
  \BibitemOpen
  \bibfield  {author} {\bibinfo {author} {\bibfnamefont {N.}~\bibnamefont
  {Sachdeva}} \emph {et~al.},\ }\bibfield  {title} {\enquote {\bibinfo {title}
  {{New Limit on the Permanent Electric Dipole Moment of $^{129}\text{Xe}$
  Using $^{3}\text{He}$ Comagnetometry and SQUID Detection}},}\ }\href
  {https://doi.org/10.1103/PhysRevLett.123.143003} {\bibfield  {journal}
  {\bibinfo  {journal} {Physical Review Letters}\ }\textbf {\bibinfo {volume}
  {123}},\ \bibinfo {pages} {143003} (\bibinfo {year} {2019})}\BibitemShut
  {NoStop}%
\bibitem [{\citenamefont {Thiel}\ \emph {et~al.}(2007)\citenamefont {Thiel},
  \citenamefont {Schnabel}, \citenamefont {Knappe-Gr{\"{u}}neberg},
  \citenamefont {Stollfu{\ss}},\ and\ \citenamefont {Burghoff}}]{Thiel2007}%
  \BibitemOpen
  \bibfield  {author} {\bibinfo {author} {\bibfnamefont {F.}~\bibnamefont
  {Thiel}}, \bibinfo {author} {\bibfnamefont {A.}~\bibnamefont {Schnabel}},
  \bibinfo {author} {\bibfnamefont {S.}~\bibnamefont {Knappe-Gr{\"{u}}neberg}},
  \bibinfo {author} {\bibfnamefont {D.}~\bibnamefont {Stollfu{\ss}}},\ and\
  \bibinfo {author} {\bibfnamefont {M.}~\bibnamefont {Burghoff}},\ }\bibfield
  {title} {\enquote {\bibinfo {title} {{Demagnetization of magnetically
  shielded rooms}},}\ }\href {https://doi.org/10.1063/1.2713433} {\bibfield
  {journal} {\bibinfo  {journal} {Review of Scientific Instruments}\ }\textbf
  {\bibinfo {volume} {78}},\ \bibinfo {pages} {035106} (\bibinfo {year}
  {2007})}\BibitemShut {NoStop}%
\end{thebibliography}%

\end{document}